\documentclass[11pt]{article}

\usepackage{geometry}
\usepackage[algo2e,ruled,lined,titlenumbered]{algorithm2e}

\newcommand{\V}[1]{\underline{#1}}
\newcommand{\M}[1]{\underline{\underline{#1}}}

\newcommand{\disp}{\V{u}}

\newcommand{\pos}{\V{x}}
\newcommand{\disptest}{\V{u}^{\star}}

\newcommand{\domega}{\partial \Omega}
\newcommand{\domegau}{\partial \Omega_u}
\newcommand{\domegaf}{\partial \Omega_f}
\newcommand{\ud}{\V{U}_\textrm{D}}
\newcommand{\fd}{\V{f}_\textrm{D}}
\newcommand{\Fd}{\V{F}_\textrm{D}}
\newcommand{\stress}{\M{\sigma}}
\newcommand{\strain}{\M{\epsilon}(\disp)}

\newcommand{\straintest}{\M{\epsilon}(\disptest)}

\newcommand{\dispdiscr}{\V{\mathbf{U}}}
\newcommand{\deltadispdiscr}{\V{\boldsymbol{\delta} \mathbf{U}}}
\newcommand{\resdiscr}{\V{\mathbf{R}}}
\newcommand{\resdiscrred}{\V{\mathbf{R}}_R}
\newcommand{\resdiscrhyperred}{\V{\mathbf{R}}_\textrm{HR}}
\newcommand{\fint}{\V{\mathbf{F}}_{\textrm{Int}}}
\newcommand{\fext}{\V{\mathbf{F}}_{\textrm{Ext}}}
\newcommand{\zerodiscr}{\V{\mathbf{0}}}

\newcommand{\A}{\M{\mathbf{A}}}
\newcommand{\B}{\V{\mathbf{B}}}

\newcommand{\proj}{\M{\mathbf{P}}}
\newcommand{ \precond}{\M{\widetilde{\mathbf{M}}}}
\newcommand{\optang}{\M{\mathbf{K}}_\textrm{T}}
\newcommand{\stiff}{\M{\mathbf{K}}}
\newcommand{\stiffred}{\M{\mathbf{K}}_\textrm{R}}
\newcommand{\optangtilde}{\M{\mathbf{\widetilde{K}}}_\textrm{T}}
\newcommand{\optanghat}{\M{\mathbf{\widehat{K}}}_\textrm{T}}
\newcommand{\optangE}{\M{\mathbf{K}}_{\textrm{T},E}}
\newcommand{\optangred}{\M{\mathbf{K}}_\textrm{T,R}}
\newcommand{\optanghyperred}{\M{\mathbf{K}}_\textrm{T,HR}}
\newcommand{\basereduc}{\M{\mathbf{C}}}
\newcommand{\coeffreduc}{\V{\boldsymbol{\alpha}}}
\newcommand{\deltacoeffreduc}{\V{\boldsymbol{\delta \alpha}}}

\newcommand{\critnew}{\nu_\textrm{New}}
\newcommand{\critnewred}{\nu_\textrm{New,R}}
\newcommand{\critCG}{\nu_\textrm{CG}}

\newcommand{\cRitz}{\M{\mathbf{c}}}
\newcommand{\Deltal}{\Delta l}

\newcommand{\resdiscreq}{\V{\mathbf{R}}_{\textrm{eq}}}
\newcommand{\resdiscrcont}{R_{\textrm{cont}}}

\usepackage{amssymb} 
\usepackage{amsmath} 
\usepackage{stmaryrd} 
\usepackage{graphicx}
\usepackage{comment}

\title{Bridging Proper Orthogonal Decomposition methods and augmented Newton-Krylov algorithms: an adaptive model order reduction for highly nonlinear mechanical problems}

\author{P. Kerfriden$^{1}$, P. Gosselet$^{2}$ \footnote{CNRS Research Associate} , S. Adhikari$^{3}$ \footnote{Chair of Aerospace Engineering} , S. Bordas$^{1}$ \footnote{Royal Academy of Engineering/Leverhulme Senior Research Fellow}
\\ \\
$^{1}$ Cardiff University \\ Queen's Buildings, The Parade, Cardiff CF24 3AA, Wales, UK \\
$^{2}$ Ecole Normale Superieure de Cachan \\ 61 Avenue du Pr\'esident Wilson, F-94230 Cachan, France \\
$^{3}$ Swansea University \\ Singleton Park, Swansea, SA2 8PP, Wales UK 
}

\begin{document}
\maketitle






\begin{abstract}
This article describes a bridge between POD-based model order reduction techniques and the classical Newton/Krylov solvers. This bridge is used to derive an efficient algorithm to correct, ``on-the-fly", the reduced order modelling of highly nonlinear problems undergoing strong topological changes. Damage initiation problems are addressed and tackle via a corrected hyperreduction method. It is shown that the relevancy of reduced order model can be significantly improved with reasonable additional costs when using this algorithm, even when strong topological changes are involved. \\

\end{abstract}



\section{Introduction}


The simulation of failure in complex materials has been one  important issue throughout engineering. 
Micro and mesoscale models have demonstrated their ability to predict complex phenomena such as crack initiation and propagation in various fields: composite materials in aeronautics \cite{lemaitrechaboche1990,lubineauladeveze2007}, 
cement-based structures \cite{jeffersonbennett2009} in civil engineering or biostructures \cite{franceschinibigoni2006} in medical engineering. 
However, these models usually describe the material at a very fine scale compared to the size of the structure, 
and the finite element discretization of the underlying partial differential equations leads to large, 
potentially highly nonlinear (therefore requiring a fine discretization in time), numerical problems. 
Various multiscale computational strategies have been developed to tackle this important issue, such as enrichment techniques \cite{moesdolbow1999,bordasmoran2006},
homogenization strategies \cite{fishshek1997,feyelchaboche2000}, domain decomposition methods \cite{gosseletrey2006,ladevezepassieux2009}, or model order reduction\cite{karhunen1947,loeve1963}. The most efficient of these techniques couple
several of these advanced tools to efficiently perform multiscale simulations  \cite{ladevezenouy2003,yvonnethe2007,pebrelrey2008,kerfridenallix2009,ladevezepassieux2009,chinestaammar2009,ryckelynckbenziane2010}.

Among them, Proper Orthogonal Decomposition-based (POD) \cite{karhunen1947,loeve1963,lianglee2002} model order reduction strategies provide extremely valuable tools 
for an automatic derivation of a multiscale computational method 
(no \textit{a priori} physical understanding of the different scales involved is required). 
They consist in using a set of potential solutions to the initial problem (snapshot) and extracting, by a spectral analysis a few basis vectors spanning a space of small dimension in which the solution to the, initially large, numerical problem is  well approximated \cite{sirovich1987}.
These techniques have been proved extremely efficient when speed is more important that accuracy, for instance if the goal is to provide an approximation of the response of a complex structure in real-time (an example of biomedical application is given in \cite{niroomandialfaro2008}), or quasi real-time (in the case of early stage design). This type of applications is our main focus.

In the context of structural problems involving plasticity or damage, two severe drawbacks limit the direct application of POD-based model order reduction:
\begin{itemize}
\item The initial snapshot might be too poor to represent accurately the solution of the damaged structure. This can happen if:
\begin{itemize}
\item The ``a priori" mechanical understanding of the structure is too poor, which might result in a bad choice of the snapshot simulations.
\item The number of parameters involved is too important. The snapshot needed to compute a relevant reduced basis might be very large, and its reduction inefficient.
\item Strong topological changes in the structure occur. Improvements have been proposed by Ryckelynk's team \cite{ryckelynck2005} where the
reduced basis is enriched during the computation. When required, a Krylov subspace associated to a linearised operator of the initial problem is generated, and a few 
additional basis vectors are obtained by a spectral analysis of this space. However, it seems that, when dealing with complex constitutive laws, 
this adaptive strategy can be computationally expensive \cite{ryckelynckbenziane2010}.
\end{itemize}
 \item The integration of the constitutive law needs to be done at each integration point, regardless of the dimension of the reduced space. These problems have been handled in \cite{ryckelynck2008} by a technique called hyperreduction, which consists in considering only a few local residuals to compute the internal forces. 
\end{itemize}

On the other hand, various model order reduction strategies have also been used to derive good initializations and preconditioners for iterative algorithms for complex nonlinear problems at the fine scale \cite{markovinovicjansen2006,rislerrey2000,gosseletrey2006,pebrelrey2008,kerfridenallix2009}. 

We propose here a novel strategy that couples these two approaches. Our vision is that if complex changes in the topology of the structure appear 
(local crack initiation for instance), they can only be accurately predicted 
by solving the fine scale model, at least locally. This can be done, for instance, by adding the finite element shape functions to the snapshot, or by using relocalization techniques. 
However, the long-range effects of these topological changes do not require the fine description, and can be obtained by solving the full system very crudely. Hence, when the  global residual exceeds a given threshold value, we propose to perform conjugate gradient iterations on the full system, orthogonally to the snapshot space by means of a classical projection. The new vector obtained by the conjugate gradient does not belong to the initial approximation space and is added to the reduced basis. It is interpreted as an``on-the-fly" enrichment of the reduced model. The proposed algorithm largely differs from what had been done in \cite{ryckelynck2005} in the sense that only a single vector belonging to the Krylov algorithm is used to enrich the reduced basis, for each of the corrections performed within a time step. It thus keeps the size of the reduced basis to a minimum. In addition, the solution of the current time step is not restarted after enriching the reduced basis. At last, using the properties of the projected Krylov during the correction steps automatically ensures the required orthogonality between the additional reduced basis vectors and the initial reduced basis.

This strategy can be interpreted as a bridge between ``exact" Newton-Krylov strategies, and projection-based model order reduction methods (like the POD) for nonlinear problems. Indeed, setting the residual threshold to a low value leads to the former strategy, while setting it to a high value leads to the latter.  
An intermediate value yields an adaptive model order reduction method with control of the global residual. We will show that this method permits to obtain a correction of the reduced model at cheap price, when the current reduced basis is no more sufficient to represent the solution accurately. 

The paper is organized as follows. In section 1, the nonlinear system of equations resulting from the discretization of a continuum mechanics problem is introduced, and the model order reduction and associated time solution procedure are described. In section 2, we perform the reduction of a damage model, and show that the basic application of the proper orthogonal decomposition is inadequate. The corrective reduced order modelling is introduced in section 3, and an exhaustive numerical study of its efficiency is proposed in section 4. In the last section, this technique is applied to the ``on-the-fly'' correction of hyperreduced damage models.

\section{Problem statement and classical model order reduction for nonlinear problems}

\subsection{Problem statement}

\subsubsection{Nonlinear structural Problem}

Let us consider a structure occupying a continuous domain $\Omega$ with boundary $\domega$. This structure is subjected to prescribed displacements $\ud$ on its boundary $\domegau$, over time interval $[0 , T]$. 
Let $\disp$ be the unknown displacement field, it belongs to the space $\mathcal{U}$ of kinematically admissible fields:
\begin{equation}
\label{eq:kinematic}
\mathcal{U}=\left\{\disp\in H^1(\Omega) \ | \  \disp_{| \domegau} = \ud\right\}
\end{equation}

Let $\mathcal{U}_0$ be the associated vector space.
Prescribed tractions $\Fd$ are applied on the complementary boundary $\domegaf = \domega \backslash \domegau$. Under the assumptions of quasi-static evolution of the structure and small perturbations, the weak form of the equilibrium and the constitutive relation read, at any time $t \in [0,T]$:
\begin{equation}
\label{eq:equilibrium}
\begin{array}{l}
\displaystyle \forall \disptest \in \mathcal{U}_0, \ \text{find} \ \disp \in \mathcal{U} \ \text{such that:}
 \\ \displaystyle \qquad \qquad
\int_{\Omega} \stress : \straintest \ d \Omega =  \int_{\Omega} \fd . \disptest \ d \Omega + \int_{\domegaf} \Fd . \disptest \ d \Gamma\\
\displaystyle \qquad \qquad \stress = \stress \left( \left( \strain_{|\tau} \right)_{\tau \leq t} \right)
\end{array}
\end{equation}
where $\stress$ is the Cauchy stress tensor and $\strain$ is the symmetric part of the gradient of displacements. 
The constitutive relation between $\stress$ and $\strain$ is nonlinear and described using internal variables (for instance damage or plasticity). It is assumed to be local and rate-independent.

\subsubsection{Space discretization}

We perform a standard finite element approximation of the space of unknown displacement field $\mathcal{U}$ (and a similar approximation of the space of test functions $\mathcal{U}_0$):
\begin{equation}
\mathcal{U}^{h}(\Omega) = \left\{ \disp(\V{x}) \ | \ \disp(\V{x}) = \sum_{i=1}^{n_n} N_i(\pos) \, \disp_i  \right\}
\end{equation}
where $n_n$ is the number of nodes, $\pos$ is the position vector, $(N_i)_{i \in \llbracket 1 \ n_n \rrbracket}$ are the finite element shape functions, and $(\disp_i)_{i \in \llbracket 1 \ n_n \rrbracket }$ the nodal values of the displacement field.

The introduction of the finite element approximation into equations \eqref{eq:kinematic} and \eqref{eq:equilibrium} leads to the following nonlinear vectorial equation at any time $t \in [0 , T]$:
\begin{equation}
\label{eq:discr_problem}
\fint \left( \left(\dispdiscr_{| \tau} \right)_{\tau \leq t} \right) + \fext=\zerodiscr
\end{equation}
where $\dispdiscr \in \mathbb{R}^{n_u}$ ($n_u$ is the number of nodal unknowns) is the vector of nodal displacement unknowns, $\fint \in \mathbb{R}^{n_u}$ and $\fext \in \mathbb{R}^{n_u}$ are the internal forces resulting from the discretization of the internal virtual work (left-hand side of \eqref{eq:equilibrium}) and the external forces resulting from the discretization of the external virtual work (right-hand side of \eqref{eq:equilibrium}), respectively.

\subsubsection{Time discretization}

The nonlinear solution strategy used here is a classical time discretization scheme for quasi-static and rate-independent problems. 
This procedure consists in finding a set of consecutive solutions at times $(t_n)_{n \in \llbracket 0 \ n_t \rrbracket}$ (see \cite{allixkerfriden2010}). Hence, the constitutive law \eqref{eq:equilibrium} is discrete in time. 
This time discretization scheme yields the following vectorial system of $n_u$ nonlinear equations at time $t_{n+1}$:
\begin{equation}
\label{eq:discretized_equilibrium}
\fint \left( \left(\dispdiscr_{| t_m} \right)_{m \in \llbracket 0 \ n+1 \rrbracket} \right) + \fext= \zerodiscr
\end{equation}
For the sake of clarity, the dependency of the internal forces with the history of the unknown fields will not be written explicitly.

\subsubsection{Nonlinear solution strategy}

At each time $t_{n+1}$ the nonlinear system of equations is solved by a Newton-Raphson algorithm. At any iteration $(i+1)$ of the algorithm, the linearisation of equation \eqref{eq:discretized_equilibrium} around $\dispdiscr^{i} $ leads to the following prediction stage: 
\begin{equation}
\label{eq:prediction_NR}
\optang^i \, \deltadispdiscr^{i+1} = - \resdiscr^{i}
\end{equation}
where we have introduced the residual vector $\resdiscr = \fext + \fint(\dispdiscr)$, and the correction of the increment of displacement $\deltadispdiscr^{i+1} = \dispdiscr^{i+1} - \dispdiscr^{i} $. This linear prediction is followed by a correction stage:
\begin{equation}
\label{eq:correction_NR}
\resdiscr^{i+1} = \fext + \fint(\dispdiscr^{i+1}) \qquad \qquad {\optang^{i+1}} = \left. \frac{\partial \fint(\dispdiscr)}{\partial \dispdiscr} \right|_{ \dispdiscr = \dispdiscr^{i+1}}
\end{equation}

\subsection{Model order reduction}

The set of equations \eqref{eq:discretized_equilibrium} can be very large if multiscale problems are simulated (see for instance \cite{bordasmoran2006,kerfridenallix2009}). Various strategies can be used to reduce the size of this system without loosing accuracy. We apply here the classical model order reduction by projection to the solution of problem \eqref{eq:discretized_equilibrium}.

\subsubsection{Reduction by projection}

The solution vector is searched for in a space of small dimension (several orders smaller than the number of finite element degrees of freedom). Let us call $\basereduc$ the matrix whose columns form a basis of this space:
\begin{equation}
\basereduc = \left(
\begin{array}{cccc}
\V{\mathbf{C}}^{1} & \V{\mathbf{C}}^{2} & ... & \V{\mathbf{C}}^{n_\textrm{C}}
\end {array}
\right)
\end{equation}
where $n_\textrm{C}$ is the dimension of the reduced space, and $(\V{\mathbf{C}}^{k})_{k \in \llbracket 1 \ n_\textrm{C} \rrbracket} \in {(\mathbb{R}^{n_u})}^{n_\textrm{C}}$ are the chosen reduced basis vectors. Applied to the reduction of problem \eqref{eq:discretized_equilibrium}, the solution field is searched for under the form:
\begin{equation}
\dispdiscr = \dispdiscr_{| t_n}+\basereduc \, \coeffreduc 
\end{equation}
The residual of equation \eqref{eq:discretized_equilibrium} is constrained to be orthogonal to a space of small dimension, which can, in theory, be different from the one spanned by $(\V{\mathbf{C}}^{k})_{k \in \llbracket 1 \ n_\textrm{C} \rrbracket}$. Let us limit our study to the Galerkin procedure, which writes the following constraint: 
\begin{equation}
  \label{eq:ortho_cond}
  \basereduc^T \, \resdiscr = \zerodiscr
\end{equation}
Hence, the reduced form of problem \eqref{eq:discretized_equilibrium} reads:
\begin{equation}
\label{eq:discr_problem_reduc}
\basereduc^T \left( \fext + \fint(\dispdiscr_{| t_n} +\basereduc \, \coeffreduc ) \right) = \zerodiscr
\end{equation}

\subsubsection{Solution procedure for the reduced nonlinear problem}
\label{sec:sol_proc_red}

At each time step of the time discretization scheme, the reduced problem \eqref{eq:discr_problem_reduc} is solved by a Newton-Raphson algorithm. The $({i+1})^{th}$ prediction stage is performed using the following set of $n_\textrm{C}$ linearised equation:
\begin{equation}
\label{eq:prediction_NR_red}
\optangred^i \, \deltacoeffreduc^{i+1} = -\resdiscrred^{i}
\end{equation}
where the residual of the reduced nonlinear problem \eqref{eq:discr_problem_reduc} is defined by $\resdiscrred = \basereduc^T \,\fext + \basereduc^T \, \fint(\dispdiscr_{| t_n}+\basereduc \, \coeffreduc)$ and the increment  $\deltacoeffreduc^{i+1} =\coeffreduc ^{i+1} - \coeffreduc^{i}$. The correction stage reads: 
\begin{equation}
\resdiscrred^{i+1} = \basereduc^T \, \fext + \basereduc^T  \, \fint(\dispdiscr_{| t_n}+\basereduc \, \coeffreduc^{i+1}) \qquad \qquad {\optangred^{i+1}} = \basereduc^T \, \left. \frac{\partial \fint(\dispdiscr_{| t_n}+\basereduc \, \coeffreduc)}{\partial \coeffreduc} \right|_{ \coeffreduc = \coeffreduc^{i+1}}
\end{equation}

\vspace{10pt}
There is a simple link between the previous solution procedure for the reduced problem \eqref{eq:discr_problem_reduc} and the systematic reduction of the prediction stages performed on the full nonlinear problem \eqref{eq:prediction_NR} since $\optangred^i$ can be expanded as
\begin{equation}
{\optangred^{i}} = \basereduc^T \, \left. \frac{\partial \fint(\dispdiscr_{| t_n}+\basereduc \, \coeffreduc)}{\partial \coeffreduc} \right|_{ \coeffreduc = \coeffreduc^{i}} =
\basereduc^T \, \left. \frac{\partial \fint(\dispdiscr)}{\partial \dispdiscr} \right|_{\dispdiscr   = \dispdiscr_{| t_n}+\basereduc \, \coeffreduc^{i}}
\left. \frac{\partial \dispdiscr}{\partial \coeffreduc} \right|_{ \coeffreduc = \coeffreduc^{i}}=\basereduc^T \, \optang^{i} \basereduc
\end{equation}
This result can be injected in equation \eqref{eq:prediction_NR_red}, which yields:
\begin{equation}
\label{eq:prediction_NR_reduc}
\basereduc^T ( \resdiscr^{i} +  \optang^{i} \, \basereduc  \, \deltacoeffreduc^{i+1} ) = \zerodiscr
\end{equation}
This last equation is a simple reduction by projection of the $(i+1)^{th}$ linear prediction of the Newton-Raphson scheme \eqref{eq:prediction_NR} used to solve the initial problem \eqref{eq:discretized_equilibrium}. In other words, the prediction stage of any Newton iteration performed on the reduced problem yields exactly the same solution as a reduction of the linear prediction stage used to solve the initial problem. 

\subsubsection{POD reduced approximation space}

The snapshot-POD is a projection-based model order reduction technique which requires the knowledge of a representative family of solutions to the global problem. 
This set of vectors is called $(\V{\mathbf{S}}^{k})_{k \in \llbracket 1 \ n_\textrm{S} \rrbracket}$ and forms an operator $\M{\mathbf{S}} = \left( \begin{array}{cccc} \V{\mathbf{S}}^1 & \V{\mathbf{S}}^2 & ... &  \V{\mathbf{S}}^{n_\textrm{S}} \end{array} \right) $. 
The aim is to find an orthonormal basis $(\V{\mathbf{C}}^{k})_{k \in \llbracket 1 \ n_\textrm{C} \rrbracket}$, of dimension $n_\textrm{C}$ smaller than $n_\textrm{S}$ such that the distance between spaces $\textrm{Im}(\mathbf{C})$ and $\textrm{Im}(\M{\mathbf{S}})$ is minimum (in the sense of the Frobenius norm). 

This problem is classically solved by computing the singular value decomposition of $\M{\mathbf{S}}$:
\begin{equation}
\M{\mathbf{S}} = \M{\mathbf{U}} \, \M{\mathbf{\Sigma}} \, \M{\mathbf{V}}^T = \sum_{k=1}^{n_\textrm{S}} \Sigma^{k} \, \V{\mathbf{U}}^{k} \,   {\V{\mathbf{V}}^{k}}^T
\end{equation}
where $ \M{\mathbf{U}} = \left( \begin{array}{cccc} \V{\mathbf{U}}^1 & \V{\mathbf{U}}^2 & ... &  \V{\mathbf{U}}^{n_u} \end{array} \right)$ and 
$ \M{\mathbf{V}} = \left( \begin{array}{cccc} {\V{\mathbf{V}}^1} & {\V{\mathbf{V}}^2} & ... &  {\V{\mathbf{V}}^{n_\textrm{S}}} \end{array} \right)$ are 
two orthonormal matrices of respective sizes $n_u$ and $n_\textrm{S}$ and the upper block of $\M{\mathbf{\Sigma}}$ is a diagonal 
matrix of positive entries $(\Sigma^{k})_{k \in \llbracket 1 \ n_\textrm{S} \rrbracket}$ (singular values), ordered decreasingly (the lower block is a null matrix). One can show that the optimum reduced snapshot basis $(\V{\mathbf{C}}^{k})_{k \in \llbracket 1 \ n_\textrm{C} \rrbracket}$ is such that:
\begin{equation}
\basereduc = 
\left( \begin{array}{cccc}
\V{\mathbf{U}}^1 & \V{\mathbf{U}}^2 & ... &  \V{\mathbf{U}}^{n_\textrm{C}}
\end{array} \right)
\end{equation}

\subsection{POD for the reduction of damage evolution problems}

\subsubsection{Damageable lattice structure}
\label{sec:lattice}
We focus on a very simple damageable lattice structure, made of bars under traction or compression (figure (\ref{fig:Reference})). Each of the elementary bars occupies the domain $\Omega_{b}$ such that
$\Omega = \bigcup_{b \in \llbracket 1 \, n_b\rrbracket } \Omega_b$, where $n_b$ is the number of bars. The direction of the bar is $\V{n}_{kl}=\V{P_k P_l}/\|\V{P_k P_l}\|$ where $P_k$ and $P_l$ are the two nodes
connected to Bar $b$. The displacement field in $\Omega_b$ is searched for under the form: 
\begin{equation}
 \disp = u(\tilde{y}) \, \V{n}_{kl}
\end{equation}
where $\tilde{y} \in [0 \ \ \lVert \V{P_k P_l} \rVert]$. We suppose that no volume force is applied on the structure, that the material of each
elementary bar is isotropic and homogeneous (constant material stiffness along the direction of the bar), 
and that the section of the bars is constant and equal to $(S_b)_{b \in \llbracket 1 \, n_b\rrbracket }$. Hence, the unknown displacement field is 
linear along the direction of the bar.

A very basic constitutive law, based on classical damage mechanics \cite{lemaitrechaboche1990}, is used to described the behaviour
of the lattice structure. The lineic strain energy of each bar reads:
\begin{equation}
\displaystyle e_d = \frac{ 1 }{ 2 } E(1-d)  S_b  \epsilon^2
\end{equation}
where $\epsilon(\disp) = u_{,\tilde{y}}(\tilde{y})$ is the strain measured in the direction of the bar, and $d$ is a damage variable which ranges from 0 (undamaged material) to 1 
(completely damaged material) and is constant within each elementary bar. Two state equations can be derived from the strain energy. 
The first one links the normal force  $N = S_b \, \V{n}_{kl}^{T} . \stress .\V{n}_{kl}$ to the strain, while the second one links a thermodynamic force $Y$ to the strain:
\begin{equation}
\displaystyle N = \frac{ \partial e_d }{ \partial \epsilon } = E(1-d)  S_b  \epsilon    \qquad \qquad  Y = -\frac{ \partial e_d }{ \partial d } =  \frac{ 1 }{ 2 } E  S_b  \epsilon^2
\end{equation}
An evolution law is defined to link the damage variable to the thermodynamic force:
\begin{equation}
d = \min \left( 1,\sup_{\tau \leq t} \left( \alpha  ( Y_{|\tau})^\beta \right) \right)
\end{equation}

The results presented in this paper are obtained with the following set of data. The sections and Young's modulus are unitary. 
The material parameters are $\beta=0.5$ and $\alpha=\sqrt{2}$. Each vertical or horizontal bar has a unitary length. 

The finite element discretization of the lattice problem is done by considering that each bar is a linear finite element. Because of the previously made assumptions on the 
loading and geometry, the finite element solution yields the ``exact'' solution to the problem.

\subsubsection{Newton/arc-length solution scheme}

\begin{figure}[htb]
       \centering
       \includegraphics[width=0.7 \linewidth]{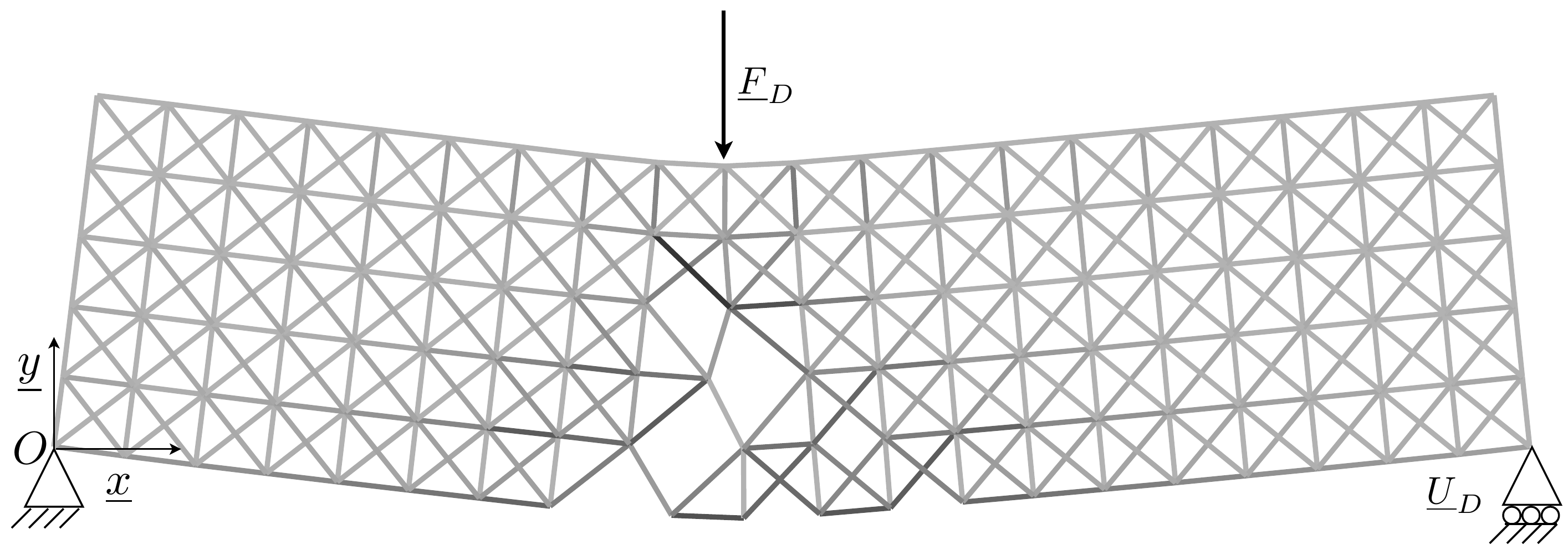}
       \caption{Reference problem. The structure is a lattice made of damageable bars in traction or compression. The problem becomes unstable as damage localizes and propagates.}
       \label{fig:Reference}
\end{figure}

The family of problems that we want to solve is described in figure (\ref{fig:Reference}). The load is applied on the top surface of the structure, in the $\underline{y}$ direction. The damage state represented here (light gray corresponds to $d=0$, while darker bars are damaged, and ruined bars are removed) is obtained in 60 time increments, each of which is converged to a very low level of error. 

This problem being unstable, a local arc-length procedure is combined to the Newton algorithm. This classical \cite{schellenkensde-borst1993,allixcorigliano1996} procedure is usually derived for a particular case of the modified Newton algorithm, namely the updated secant Newton. In our work, we need a certain versatility to choose the linearised operator. Therefore, the local arc-length procedure is detailed here for tangent and modified Newton algorithms, in a general manner.

At time $t_{n+1}$, this procedure simply consists in relaxing the norm $\lambda$ of the prescribed loading,
while adding a local constraint on the increments of displacement:
\begin{equation}
\label{eq:SysRitz}
 \left\{ \begin{array}{l}
          \displaystyle  \fint \left( \left(\dispdiscr_{| t_m} \right)_{m \in \llbracket 0 \ n+1 \rrbracket} \right) + \lambda \, \fext = \zerodiscr
	  \\
	  \displaystyle  \cRitz \left( \left(\dispdiscr_{| t_m} \right)_{m \in \llbracket 0 \ n+1 \rrbracket} \right) \Delta \dispdiscr = \Deltal 
\end{array} \right.
\end{equation}
where $\fext$ is normalized, and $\Delta \dispdiscr = \dispdiscr_{|t=t_{n+1}}-\dispdiscr_{|t=t_n}$. $\cRitz$ is a formal boolean line operator which extracts the maximum difference of displacement increment undertaken by an element (whose stiffness is strictly positive) over the current time step:
\begin{equation}
\label{eq:control_eq}
\cRitz \left( \left(\dispdiscr_{| t_m} \right)_{m \in \llbracket 0 \ n+1 \rrbracket} \right) \Delta \dispdiscr =\max_{ \small \begin{array}{cc} b \in \llbracket 1 \, n_b\rrbracket \\ \forall M \in \Omega_b, \, d|_{M,t=t_n} < 1 \end{array} } \left(   \Delta \dispdiscr |_{P_l} - \Delta \dispdiscr |_{P_k} \right) . \V{n}_{kl}
\end{equation}
where $\Delta \dispdiscr|_{P_k}$ and $\Delta \dispdiscr |_{P_l}$ are the local increments of displacements over the 
time interval $[t_n , t_{n+1}]$, respectively at corners $P_k$ and $P_l$ of the bar.

These two equations are linearised around point $(\lambda^i,\dispdiscr^i)$, which yields, at step $(i+1)$ of the Newton algorithm, the following linear prediction:
\begin{equation} 
 \left\{ \begin{array}{l}
\displaystyle \delta \lambda^{i+1} = \frac{-\resdiscrcont^{i} - \cRitz^{i} \, (\optang^{i})^{-1} \resdiscreq^i  }  {\displaystyle \cRitz^{i} \, ({\optang^{i}})^{-1} \fext}
\\
\displaystyle \deltadispdiscr^{i+1} = - ({\optang^{i}})^{-1} \left( \resdiscreq^i + \delta \lambda^{i+1} \, \fext \right)
\end{array} \right.
\end{equation}
with the definition of the residuals of system \eqref{eq:SysRitz}, $\resdiscreq = \fint + \lambda \, \fext  $, $\resdiscrcont = \Deltal -\cRitz \, \Delta \dispdiscr $ and the increment $\delta \lambda^{i+1} =\lambda^{i+1} - \lambda^{i}$.
During the correction stage, the tangent stiffness operator and residuals are updated similarly as described for the classical Newton algorithm, equation \eqref{eq:correction_NR}, while the new extraction operator is corrected as follows:
\begin{equation} 
\begin{array}{l}
\cRitz^{i+1} = \cRitz \left( \dispdiscr^{i+1} , \left(\dispdiscr_{| t_m} \right)_{m \in \llbracket 0 \ n \rrbracket} \right) 
\end{array}
\end{equation}

As suggested in \cite{allixcorigliano1996,kerfridenallix2009}, we use a quasi-Newton approximation of the previously described tangent Newton-Raphson algorithm. 
The tangent operator at iteration ${\optang^{i}}$ is replaced by the stiffness matrix $\stiff^{i}$ obtained by setting the internal variables to previously obtained values (i.e.: 
last converged solution, or last Newton iteration). This procedure does not yield the quadratic convergence rate of the genuine Newton-Raphson algorithm. Yet, it is widely
 applied in damage simulations for its simplicity (no need to compute a tangent operator) and robustness. 

The same arc-length procedure is of course applied when reducing the nonlinear problem by projection on a reduced basis basis. The expressions of 
the tangent stiffness and residual are similar to the one given in section \eqref{sec:sol_proc_red} (the residual is slightly modified 
by the introduction of the loading parameter $\lambda$). The linear prediction performed on the reduced problem reads:
\begin{equation} 
 \left\{ \begin{array}{l}
\displaystyle \delta \lambda^{i+1} = \frac{-\resdiscrcont^{i} - \cRitz^{i} \, \basereduc (\optangred^{i})^{-1} \resdiscrred^i  }  {\displaystyle \, \cRitz^{i} \, \basereduc ({\optangred^{i}})^{-1} ( \basereduc^T \, \fext ) }
\\
\displaystyle \deltacoeffreduc^{i+1} = - ({\optangred^{i}})^{-1} \left( \resdiscrred^i + \delta \lambda^{i+1} \, \basereduc^T \, \fext \right)
\end{array} \right.
\end{equation}

\subsubsection{Model order reduction}

\begin{figure}[htb]
       \centering
       \includegraphics[width=1. \linewidth]{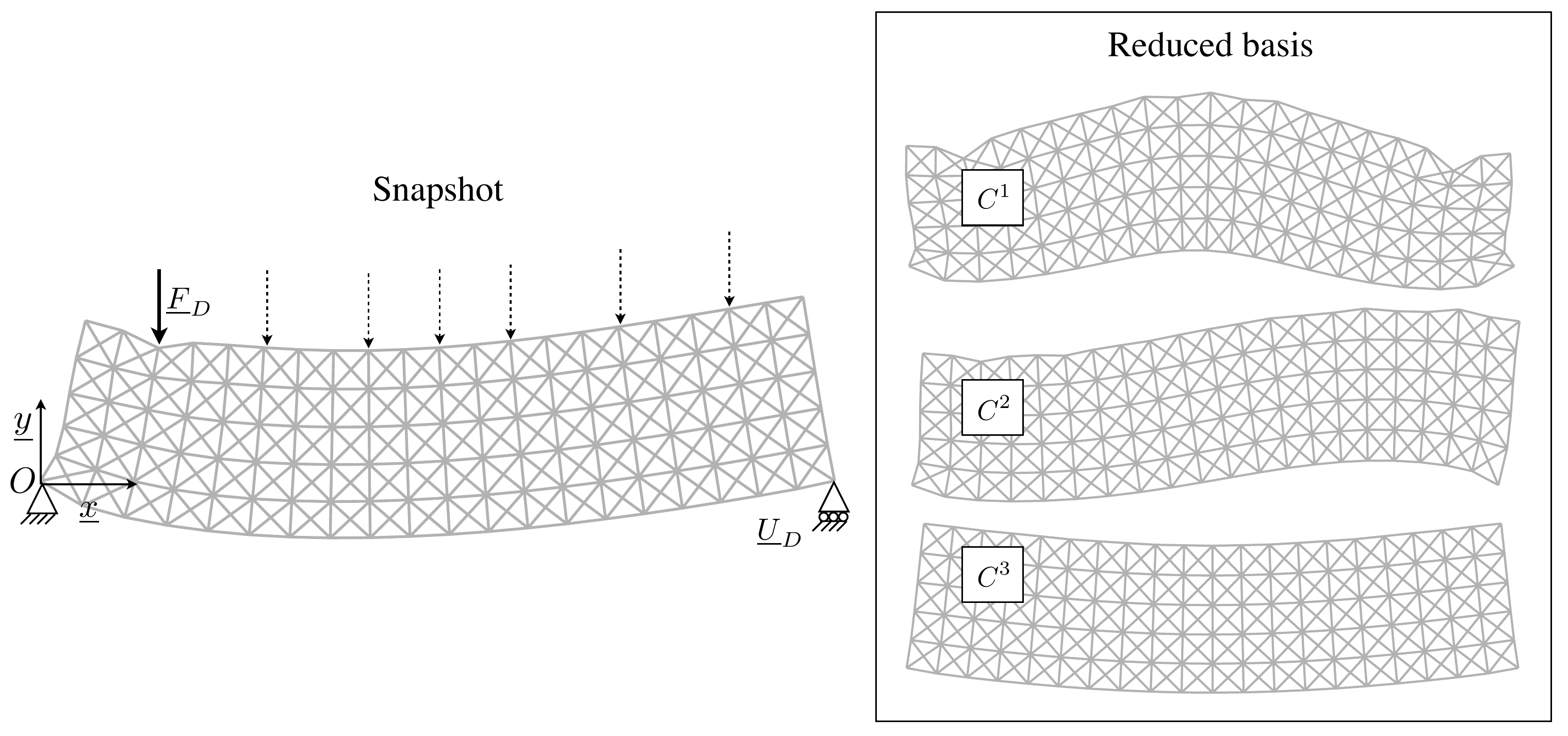}
       \caption{Computation of the snapshot made of seven solutions obtained at the very beginning of the damage process (left) and reduced basis obtained by extracting the singular vectors associated with the three highest singular values of the resulting snapshot operator (right)}
       \label{fig:Snapshot}
\end{figure}

In order to reduce the family of problems described in the previous sections by a POD-based projection method, we compute a snapshot of 7 solutions (loads applied on the top surface of the lattice, 
at positions $x \in \{ 2 \ 5 \ 8 \ 10 \ 12 \ 15 \ 18 \} $), as illustrated in figure (\ref{fig:Snapshot}). 
The successive loads have a very low value, so that the structure is virtually undamaged. This snapshot is reduced to three basis vectors by the computation of a truncated SVD.

\begin{figure}[htb]
       \centering
       \includegraphics[width=0.72 \linewidth]{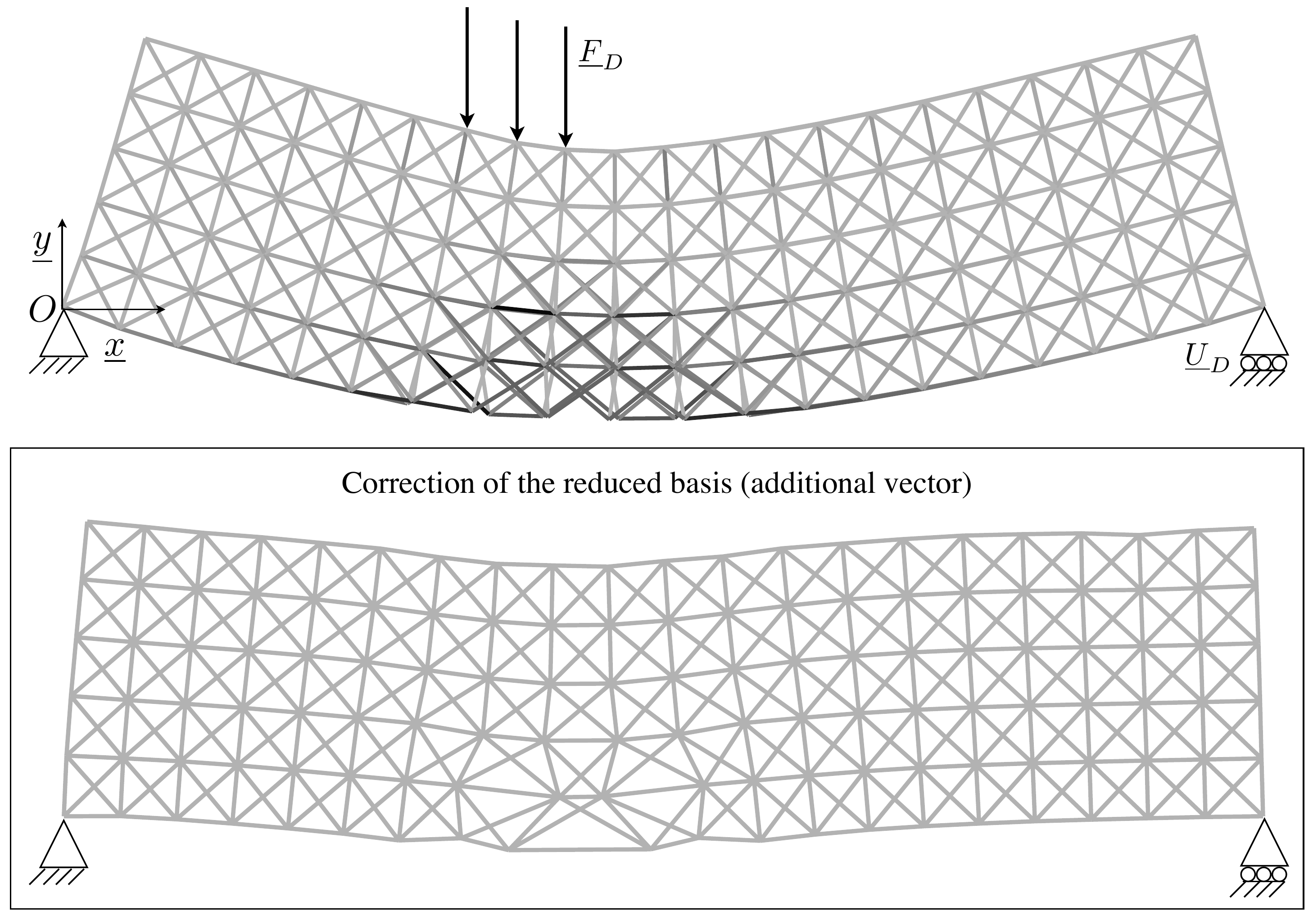}
       \caption{Superposed reference solution and solution obtained with the C-POD, for $\critnew=4.10^{-1}$ (top), additional reduced basis vectors obtained as a correction of the reduced basis (bottom).}
       \label{fig:Enriched_POD_reference}
\end{figure}

Within the family considered to construct the snapshot, the particular problem which will be simulated in the following is represented in figure (\ref{fig:Enriched_POD_reference}). The vertical load is applied at points $x \in \{ 7 \ 8 \ 9  \} $, and the time of the analysis is discretized in 50 time steps. The reference solution is converged to a very low value of the norm of the residual of the Newton/arc-length algorithm (norm of the residual scaled by the norm of the right-hand side term equal to $10^{-6}$).

Given the normalized current solution $\widetilde{\V{\mathbf{U}}}_{|t} = \frac{  \dispdiscr_{|t} }{\lVert \dispdiscr_{|t} \rVert_2}$ and normalized reference solution ${\widetilde{\V{\mathbf{U}}}_{\textrm{ref}_{|t}}}= \frac{ {\dispdiscr_\textrm{ref}}_{|t} }{\lVert {\dispdiscr_\textrm{ref}}_{|t} \rVert_2}$, the solution error used to assess the accuracy of the current solution is:
\begin{equation}
\label{eq:error}
 \textrm{solution error}_{|t} =  \left\lVert  \widetilde{\V{\mathbf{U}}}_{|t} - {\widetilde{\V{\mathbf{U}}}_{\textrm{ref}_{|t}}} \right\rVert_2
\end{equation}
In the following developments, a single value can be provided for this error. It is the maximum over time of each of these errors.

In the very early stages of the simulation, the error is very low, 
which can be seen in figure (\ref{fig:Error_New_CG}, right) (the dashed line is the error defined by equation \eqref{eq:error} as a function of time, while the dotted one is the value of the maximum damage in the structure). 
However, as damage increases, which eventually leads to the cracked area represented in figure (\ref{fig:Enriched_POD_reference}), a simple linear combination of the modes of the reduced snapshot is obviously not sufficient to reproduce accurately the solution, which results in a high level of error.



\section{Error-controlled adaptive model order reduction}


\subsection{Principle}
\label{sec:error_controlled}

In order to handle complex changes in the topology of the structure, we propose to enrich the reduced basis $\mathbf{C}$, ``on-the-fly'', during the Newton iterations performed on the reduced problem \eqref{eq:discr_problem_reduc}. Prior to the actual description of the algorithm, let us remark that, at each iteration of the Newton algorithm, one can compute:
\begin{itemize}
\item the relative norm of the residual of the reduced problem \eqref{eq:discr_problem_reduc}, i.e.: $\displaystyle \frac{ \lVert \resdiscrred \rVert_2 }{ \lVert \M{\mathbf{C}}^T \fext \rVert_2}$
\item  the relative norm of the residual of the initial problem \eqref{eq:discretized_equilibrium}, i.e.: $\displaystyle \frac{ \lVert \resdiscr \rVert_2  }{ \lVert \fext \rVert_2 }$ 
\end{itemize}
Our aim is to control both these values during the solution process. More precisely, we will proceed to the next time increment only if the former is lower 
than $\critnewred$ and the latter lower than $\critnew$, with $\critnewred \ll \critnew$. Typically, $\critnewred$ is set to $10^{-6}$ in our tests, 
while the value of $\critnew$ (which, in our tests, ranges from $10^{-1}$ to $8.10^{-1}$ ) will determine the accuracy of the successive Newton solutions, 
and is an important parameter of the strategy.

The following algorithm is applied. A sufficient number of Newton iterations are performed on the reduced problem, so that 
the norm of its residual is small enough ($k_{\textrm{Res}}$ times smaller than the norm of the residual of the full problem, with $k_{\textrm{Res}}>10$
a parameter of the method which can be tuned depending on the application and implementation). 
If the norm of the residual of the full problem is still higher than a desired value $\critnew$, the following linear 
prediction is enhanced by performing a few iterations of Conjugate Gradient on the linear prediction of the full problem \eqref{eq:discretized_equilibrium}.
The resulting solution is eventually added to the reduced snapshot basis, as detailed in the next sections.

This process is repeated, until convergence of both the reduced problem ($\lVert \resdiscrred \rVert_2 / \lVert \M{\mathbf{C}}^T \fext \rVert_2 \leq \critnewred$), 
and the full problem ($\lVert \resdiscr \rVert_2  / \lVert \fext \rVert_2 \leq \critnew$).

\subsection{Corrections on the linear predictions by an iterative solver}

When required by the previously described algorithm, the enhanced linear prediction is performed by initializing a projected conjugate gradient and performing 
a few iterations on the linearised problem \eqref{eq:prediction_NR}. The stopping criterion on the normalized residual is denoted by $\critCG$ and set to a high value, 
typically of the same order than $\critnew$. The reasons for choosing this iterative algorithm are the following:
\begin{itemize}
\item The construction of a Krylov basis requires only matrix/product operations. Therefore, no matrix assembly is required, which makes it particularly suitable for parallel computing, and ensures the versatility of  our method. In the case of serial computing, no matrix factorization is needed, which permits to save (i) memory, as we keep the sparsity of the operator (ii) time if only a coarse solution is needed, which is the case in the following developments.
\item The projection framework is an ideal tool for the particular issue of enriching the reduced basis generated by the POD. Indeed, the initialization of the projected 
algorithm is the linear prediction of the reduced problem, equation \eqref{eq:prediction_NR_red}. 
The complementary part of the solution is searched for is a space orthogonal to the reduced basis, 
which extends immediately the solution space and  reduces the number of iterations required to reach $\critCG$ (left preconditioning). 
To summarize these ideas, each iteration of the CG is a correction of the solution to the reduced linear prediction \eqref{eq:prediction_NR_red} such 
that this correction is orthogonal to the initialization.
\end{itemize}

\subsubsection{Projected Krylov solver}

The classical projected (or augmented) conjugate gradient \cite{dostal1988} is applied to the approximate solution of \eqref{eq:prediction_NR} (the linearised operator is assumed symmetric, positive and definite), which is recalled here: 
\begin{equation}
\optang \, \deltadispdiscr = \V{\mathbf{F}}
\end{equation}
Where the $i$ superscripts indicating the iteration number have been dropped for the sake of clarity, and $\V{\mathbf{F}}=-\V{\mathbf{R}}$. The chosen augmentation space is the previously defined reduced snapshot space $\textrm{Im} (\basereduc)$.

The fundamental idea of the augmented Krylov algorithms is to split the search space into two supplementary spaces:
\begin{equation}
\mathbb{R}^{n_u} = \textrm{Im} (\basereduc)  \oplus \textrm{Im} (\basereduc)^{\bot_{\optang}} 
\end{equation}
$\bot_{\optang}$ designing the $\optang-$orthogonality. The unknown solution vector is decomposed in this space under the form:
\begin{equation}
\displaystyle \deltadispdiscr = \deltadispdiscr_\textrm{C} + \deltadispdiscr_\textrm{K} \quad \textrm{where} \quad
\left\{ \begin{array}{c}
\displaystyle \deltadispdiscr_\textrm{C}=\basereduc \, \deltacoeffreduc \in \textrm{Im}(\basereduc) \\ 
\displaystyle \deltadispdiscr_\textrm{K} \in \textrm{Im} (\basereduc)^{\bot_{\optang}} = \textrm{Ker}(\basereduc^T \optang)
\end{array} \right.
\end{equation}
The $\optang-$orthogonality is ensured by introducing a projector $\proj$ such that:
\begin{equation}
\left\{ \begin{array}{l}
 \deltadispdiscr_\textrm{K} = \proj \, \mathbf{\deltadispdiscr}_\textrm{K}
\\
\basereduc^T \, \optang \, \proj = \zerodiscr
\end{array} \right.
\end{equation}
Hence, the decomposition of the correction of the displacement increment reads:
\begin{equation}
\deltadispdiscr = \basereduc \, \deltacoeffreduc + \proj \, \mathbf{\deltadispdiscr}_\textrm{K}
\end{equation}
with the classical projector:
\begin{equation}
\proj = \M{\mathbf{I}}_\textrm{d} - \basereduc (\basereduc^T \optang \basereduc  )^{-1} \basereduc^T \optang
\end{equation}

This separation of the search space into two subspaces $\textrm{Im}(\basereduc)$ and $\textrm{Im}(\proj)$ in direct sum leads to the following uncoupled equations:
\begin{equation}
\label{eq:line_ini}
(\basereduc^T \, \optang \, \basereduc) \, \deltacoeffreduc = \basereduc^T \, \V{\mathbf{F}}
\quad \Rightarrow\quad\deltadispdiscr_\textrm{C}= \basereduc(\basereduc^T \, \optang \, \basereduc)^{-1}\basereduc^T \,  \V{\mathbf{F}}
\end{equation}
\begin{equation}
\label{eq:line_proj}
\left( \optang \, \proj \right)  \mathbf{\deltadispdiscr}_\textrm{K} =  \V{\mathbf{F}} - \optang \, \deltadispdiscr_\textrm{C}
\end{equation}
Note that $\V{\mathbf{R}}_\textrm{C} =  \V{\mathbf{F}} - \optang \, \deltadispdiscr_\textrm{C}=\proj^T \,  \V{\mathbf{F}}$ and $\optang \, \proj= \proj^T \, \optang \, \proj $.
 Equation \eqref{eq:line_ini} is a coarse initialization of the projected conjugate gradient. As stated previously, it is equivalent to the linear prediction of the reduced problem (equation \eqref{eq:prediction_NR_red}). Equation \eqref{eq:line_proj} is the linear prediction of the full problem projected on $\textrm{Im} (\basereduc)^{\bot_{\optang}}$. This system is symmetric and can be solved by a preconditioned conjugate gradient. Hence we solve:
\begin{equation}
\precond^{-1} \left( \proj^T \, \optang  \, \proj \right) \deltadispdiscr_K = \precond^{-1} \, \proj^T \, \V{\mathbf{R}}_\textrm{C}
\end{equation}
where  $ \precond^{-1} $ is a left preconditioner (symmetric, definite and positive). In our test cases, $\precond$ is a diagonal matrix whose entries are the elements of the diagonal of $\optang$. Algorithm~\ref{APCG} describes the augmented preconditioned conjugate gradient.

\begin{algorithm2e}[ht]\caption{Augmented preconditioned conjugate gradient}\label{APCG}
Compute $\optang \basereduc$, $(\basereduc^T \optang  \basereduc)^{-1}$ \quad; \quad($\proj=\M{\mathbf{I}}_\textrm{d} - \basereduc\left(\basereduc^T\optang \basereduc\right)^{-1}\basereduc^T \optang$)\;  %
$\deltadispdiscr_\textrm{C} = \basereduc(\basereduc^T \optang  \basereduc)^{-1}\basereduc^T  \V{\mathbf{F}}$ \quad and\quad  ${\mathbf{\deltadispdiscr}_\textrm{K}}_0=0$\;%
$\V{\mathbf{R}}_0=\V{\mathbf{R}}-\optang \deltadispdiscr_\textrm{C}$\;%
$\V{\mathbf{Z}}_0 = \proj\precond^{-1} \V{\mathbf{R}}_0$, $\V{\mathbf{W}}_0=\V{\mathbf{Z}}_0$\;%
\For{$j=1,\ldots,n$}{%
  $\alpha_{j-1}=(\V{\mathbf{R}}_{j-1},\V{\mathbf{W}}_{j-1})/(\optang  \V{\mathbf{W}}_{j-1},\V{\mathbf{W}}_{j-1})$  \\
  $ {\mathbf{\deltadispdiscr}_\textrm{K}}_{j}= {\mathbf{\deltadispdiscr}_\textrm{K}}_{j-1}+\alpha_{j-1} \V{\mathbf{W}}_{j-1}$\\
  $\V{\mathbf{R}}_{j}=\V{\mathbf{R}}_{j-1}-\alpha_{j-1} \optang  \V{\mathbf{W}}_{j-1}$\\
  $\V{\mathbf{Z}}_{j} = \proj \precond^{-1} \V{\mathbf{R}}_{j+1}$\\
  $\beta_j=(\optang \V{\mathbf{W}}_{j-1},\V{\mathbf{Z}}_{j})/(\V{\mathbf{W}}_{j-1},\optang  \V{\mathbf{W}}_{j-1})$ \\
  $\V{\mathbf{W}}_{j}=\V{\mathbf{Z}}_{j} - \beta_j \V{\mathbf{W}}_{j-1}$

}%
\end{algorithm2e}

\subsubsection{Adaptation of the reduced basis}

The augmentation space, also used as reduced space for the projection of the linear prediction performed 
on the reduced problem, is chosen as:
\begin{equation}
\basereduc = 
\left( \begin{array}{ccc}
\basereduc_{\textrm{POD}} & \basereduc_{\textrm{NewT}} & \basereduc_{\textrm{CG}} 
\end{array} \right)
\end{equation}
where:
\begin{itemize}
 \item $\basereduc_{\textrm{POD}}$ is the initial reduced snapshot (classical POD). It is not changed, unless the size of the adaptive reduced basis becomes too large. This issue is discussed in the following.

 \item $\basereduc_{\textrm{NewT}}$ is a set of converged solution vectors obtained during previous time steps, 
when at least one enhanced linear prediction has been performed (otherwise the solution obtained is a simple linear combination of the reduced basis vectors). 
At the end of such a time step $t_n$, the solution ${\dispdiscr}_{t=t_n}$ is orthonormalized with respect to $\basereduc_{\textrm{POD}}$ and $\basereduc_{\textrm{NewT}}$
by a Gram-Schmidt procedure (using the classical Euclidian dot product). 
This orthonormalized vector is denoted by ${\widetilde \dispdiscr}_{t=t_n}$ and used to adapt $\basereduc$:
\begin{equation}
 {\basereduc_{\textrm{NewT}}}_{|t=t_{n+1}} = 
\left( \begin{array}{cc}
 {\basereduc_{\textrm{NewT}}}_{|t=t_n} & {\widetilde \dispdiscr}_{t=t_n}
\end{array} \right)
\end{equation}

 \item $\basereduc_{\textrm{CG}}$ is a set of solution vectors obtained during the Newton process at the current time step. let
$i+1$ be the current Newton iteration, the linear prediction requiring an enrichment by the augmented Conjugate Gradient.
If ${\bar \deltadispdiscr}^{i+1}$ is the solution obtained at the end of this projected iterative linear solver, it is added to
the reduced basis:
\begin{equation}
 {\basereduc_{\textrm{CG}}}^{i+1} = 
\left( \begin{array}{cc}
 {\basereduc_{\textrm{CG}}}^{i} & {\bar \deltadispdiscr}^{i+1}- \deltadispdiscr_\textrm{C}
\end{array} \right)
\end{equation}
where $\deltadispdiscr_\textrm{C}$ is the initialization of the projected iterative solver, 
given by solving the first row of system \eqref{eq:line_ini}. 
The linearised reduced operator in equation \eqref{eq:prediction_NR_red} is recomputed, 
and the linear prediction \eqref{eq:prediction_NR_red}
is performed, which yields the new solution vector ${\dispdiscr}^{i+1}$. Note that ,by construction, $( {\bar \deltadispdiscr}^{i+1} - \deltadispdiscr_\textrm{C} )$
is ${\optang}^{i}-$orthogonal to the previous reduced basis ${\basereduc}^{i}$.

When the convergence of the current time step is achieved, $\basereduc_{\textrm{CG}}$ is emptied.

\end{itemize}

The size of the reduced basis needs to be controlled. Indeed, the size of $\basereduc_{\textrm{NewT}}$ increases during the simulation. Several techniques can be used when the number of additional vectors exceeds $n_\textrm{C,NewT,Max}$:
\begin{itemize}
 \item Keep only the last vectors of $\basereduc_{\textrm{NewT}}$.
 \item Keep a few linear combination of the vectors of $\basereduc_{\textrm{NewT}}$, these linear combination being extracted by a SVD computation.
 \item Consider the following operator as a Snapshot:
\begin{equation}
\widetilde{\M{\mathbf{S}}} = 
\left( \begin{array}{cc}
{\basereduc_\textrm{POD}}_{|t=t_{n}} & {{\widetilde{\basereduc}}_{\textrm{NewT}_{|t=t_{n}}}}
\end{array} \right)
\end{equation}
where ${\widetilde{\basereduc}}_\textrm{NewT}$ is the set of all the previsous converged solutions, and choose as the new reduced basis the 
vectors associated to the highest singular values of $\widetilde{\M{\mathbf{S}}}$, in decreasing order. This new basis is split as follows:
\begin{equation}
\basereduc_{|t=t_{n+1}} = 
\left( \begin{array}{cc}
{\basereduc_{\textrm{POD}}}_{|t=t_{n+1}} & {\basereduc_{\textrm{NewT}}}_{|t=t_{n+1}}
\end{array} \right)
\end{equation}
where, ${\basereduc_{\textrm{POD}}}_{|t=t_{n+1}}$ and ${\basereduc_{\textrm{POD}}}_{|t=t_{n}}$  have the same number of columns, and
${\basereduc_{\textrm{NewT}}}_{|t=t_{n+1}}$ have a number of columns lower than the one of ${\basereduc_{\textrm{NewT}}}_{|t=t_{n}}$ (divided by two in the following examples). ${\basereduc_{\textrm{POD}}}_{|t=t_{n+1}}$
usually contains the information of the initial reduced snapshot (strongly uncoupled and associated with singular values greater than 1).
\end{itemize}
The latter technique is the one used in our examples. It has the advantage to yield a basis whose vectors have a very weak energetic coupling, hence granting the method stability (we have observed that the two former ideas, despite the orthogonality, between the vectors of the reduced basis,
easily lead to the stagnation of the Newton process when instabilities and multiple solutions have to be handled).

\begin{algorithm2e}[p]
\caption{Corrective model order reduction strategy (the control of the size of the reduced basis is not taken described)}\label{alg1} 
Load the reduced basis $\basereduc$ \;
Initialise the solution of the reduced problem \;
\ \qquad $\coeffreduc^0 = \zerodiscr$\;
\For{$n=1 .. n_t$ (time loop)}{
  Construct the linearized reduced operator \\
  \  \qquad ${\optangred^{i+1}} = \basereduc^T \, \left. \frac{\partial \fint(\dispdiscr_{n-1}+\basereduc \, \coeffreduc)}{\partial \coeffreduc} \right|_{ \coeffreduc = \coeffreduc^{0}}$\\
  Compute the initial residuals \\
  \  \qquad $ \resdiscr^{0} = \fext^n + \fint( \dispdiscr_{n-1}+\basereduc \coeffreduc^0)$ \\
  \  \qquad $\resdiscrred^{0} = \basereduc^T \resdiscr^{0} $\\
  Save the reduced basis \\
  \ \qquad  $\basereduc^0 \leftarrow \basereduc$
   \While{$ \frac{\lVert \resdiscrred \rVert_2 }{ \lVert \M{\mathbf{C}}^T \fext^n \rVert_2} \geq \critnewred$ \textrm{ and } $ \frac{\lVert \resdiscr \rVert_2  }{ \lVert \fext^n \rVert_2 } \geq \critnew$  (Newton loop)}{
     \If{$ \frac{\lVert \resdiscrred \rVert_2 }{ \lVert \M{\mathbf{C}}^T \fext^n \rVert_2}  \leq k_\textrm{Res} \frac{\lVert \resdiscr \rVert_2  }{ \lVert \fext^n \rVert_2 } $}{
       Assemble the global stiffness \\
       \ \qquad $\optang = \left. \frac{\partial \fint(\dispdiscr)}{\partial\dispdiscr} \right|_{\dispdiscr = \dispdiscr_{n-1}+\basereduc^0 \, \coeffreduc^0}$ \\
       Save the reduced basis \\
       \ \qquad  $\basereduc_{\textrm{Tmp}} \leftarrow \basereduc$ \\
       Compute the reduced basis correction by the projected conjugate gradient \\
       \ \qquad $\widetilde{\deltadispdiscr}  \approx - \optang^{-1} \resdiscr^i $ \quad \textrm{where} \quad $\widetilde{\deltadispdiscr} = \basereduc \, \widetilde{\coeffreduc} + \widetilde{\deltadispdiscr}_\textrm{K}$\\
       Update the reduced basis \\
      \ \qquad $  \basereduc \leftarrow  \left( \begin{array}{cc} \basereduc & \frac{\widetilde{\deltadispdiscr}_\textrm{K}}{ \lVert \widetilde{\deltadispdiscr}_\textrm{K} \rVert_2} \end{array} \right) $\\
       Project the previous solution in the new reduced space \\
       \ \qquad $\coeffreduc^{i} \leftarrow (\basereduc^T \basereduc)^{-1} \left( \basereduc^T \basereduc_{\textrm{Tmp}}  \, \coeffreduc^{i} \right)$
    }
    Solve the reduced linearized system \\
    \ \qquad $\deltacoeffreduc^{i+1} = - (\stiffred^i)^{-1} \resdiscrred^i$\\
    Update the solution \\ 
    \ \qquad $\coeffreduc^{i+1} = \deltacoeffreduc^{i+1} + \coeffreduc^{i}$\\
    Compute the residuals \\
    \      \qquad $ \resdiscr^{i+1} = \fext^n + \fint(\dispdiscr_{n-1}+ \basereduc \, \coeffreduc^{i+1})$ \\
    \      \qquad $\resdiscrred^{i+1} = \basereduc^T \resdiscr^{i+1} $\\ 
    $i \leftarrow i+1$
  }
  Update the reduced basis if a correction has been performed \;
  \ \qquad $  \basereduc \leftarrow \left( \begin{array}{cc} \basereduc^0 & {\widetilde \dispdiscr} \end{array} \right) $ \;
  Initialise the solution of the reduced problem for the next time step \;
  \  \qquad $\coeffreduc^0 = \coeffreduc^i$
}
\end{algorithm2e}

\section{Behaviour of the adaptative strategy}

We have introduced several parameters in the corrective model order reduction strategy (denoted by C-POD). The aim of this section is to provide an understanding of the sensitivity of the solver to a variation of these parameters, and 
when possible, to provide optimal or practical values. This study is not required to understand the method, and the reader might want to refer directly to the results given in section \ref{sec:First_results}.

\subsection{Error criteria}

In a first set of numerical experiments, we investigate the influence of the precision of the conjugate gradient iterations on the efficiency and accuracy of the method. We use the following set of fixed criteria: $k_\textrm{Res} = 10^{3}$, $n_\textrm{C,NewT,Max}=3$. The relative norm of the residual of the full problem is successively set to (a) $\critnew=4.10^{-1}$ and (b) $\critnew=5.10^{-2}$. The results obtained when decreasing threshold values $\critCG$ are reported in table 
(\ref{table:influenceCG}), Experiment A. The following tendencies can be observed:
\begin{itemize}
\item If $\critCG>\critnew$ (i.e.: the conjugate gradients are solved finely compared to the actual maximum level of error required for the convergence of the adaptive strategy), the accuracy of the algorithm (i.e: the maximum value over time of the error in solution) increases with decreasing values of $\critCG$.
\item Unless $\critCG>\critnew$ (i.e.: if the conjugate gradients are solved very loosely), the value of  $\critnew$ has no influence on the accuracy of the algorithm.
\item The number of enhanced predictions performed to reach a given accuracy is relatively independent on the accuracy of the conjugate gradients resolutions, hence independent on $\critCG$.
\item The number of cumulated iterations of the conjugate gradients increases when the accuracy of the conjugate gradients resolutions increases (i.e.: $\critCG$ decreases), which results in the global computation costs to increase.
\end{itemize}
As we want to ensure a certain consistency in the accuracy of the algorithm with respect to the reduced model error criterion $\critnew$, we set $\critCG=\critnew$. Note here that the corrective algorithm converges even if the convergence threshold on the conjugate gradients, $\critCG$, is set to a loose value.
 \\

In a second numerical experiment, we try to explain the role of $k_\textrm{Res}$ (parameter which sets the maximum value of the reduced residual norm to reach before allowing a correction to be performed) on the behaviour of the method. The following set of parameters is used, $\critnew=1.10^{-1}$, $\critCG=\critnew$,  $n_\textrm{C,NewT,Max}=3$ and the results in terms of efficiency and accuracy with different values of $k_\textrm{Res}$ are reported in table (\ref{table:influenceCG}), Experiment B. The following remarks can be made:
\begin{itemize}
\item The accuracy of the solution is not influenced by $k_\textrm{Res}$.
\item The global number of Newton iterations increases when $k_\textrm{Res}$ increases, thus increasing the global computation time.
\end{itemize}
This behaviour can be explained by a qualitative analysis of the influence of $k_\textrm{Res}$ on the solution strategy. During a time step where a correction of the reduced basis is required, using low values of $k_\textrm{Res}$ results in the corrective linear predictions to be performed at the beginning of the Newton process. Conversely, when $k_\textrm{Res}$ is set to a high value, a first nonlinear prediction is performed with the previous reduced model, and the correction is done at convergence of this process. As suggested by the results presented here, the former strategy is more efficient, as a relevant (i.e.: corrected) reduced model is constructed during the very first iterations of the Newton process.

However, we will see, in section \ref{sec:hyperred}, that this conclusion can be jeopardized by the actual implementation of the strategy.

\begin{table}[ht]
\begin{center}
  \begin{tabular}{ | c || c | c | c | c |c |  }
    \hline

     \multicolumn{6}{|c|}{EXPERIMENT A (a)} \\
     \hline 
     $\critCG$ & \textbf{CG} & \textbf{CG}  & \textbf{Newton} & \textbf{CPU} & \textbf{Solution} \\
     & \textbf{corrections} & \textbf{iterations}  & \textbf{iterations} & \textbf{time} & \textbf{error} \\ 
    \hline 
    $8.10^{-1}$ &  6 & 82  & 257 & 20 & 1.75 \% \\ 
    \hline
    $3.10^{-1}$ & 8 & 180 & 291 & 23 & 1.93 \% \\ 
    \hline
    $1.10^{-1}$ & 8 & 224 & 290 & 23 & 1.93 \% \\ 
    \hline
    $3.10^{-2}$ & 8 & 256 & 291 & 23 & 1.93 \% \\ 
    \hline
    $1.10^{-2}$ & 8 & 294 & 291 & 23 & 1.93 \% \\ 
    \hline
    $1.10^{-3}$ & 8 & 318 & 291 & 23 & 1.93 \% \\ 
    \hline

     \multicolumn{6}{|c|}{EXPERIMENT A (b)} \\
    \hline
     $\critCG$  & \textbf{CG} & \textbf{CG}  & \textbf{Newton} & \textbf{CPU} & \textbf{Solution} \\
     & \textbf{corrections} & \textbf{iterations}  & \textbf{iterations} & \textbf{time} & \textbf{error} \\ 
    \hline 
    $8.10^{-1}$ & 42 & 545  & 485 & 38 & 0.449 \% \\ 
    \hline
    $3.10^{-1}$ & 41 & 858 & 496 & 41 & 0.382 \% \\ 
    \hline
    $1.10^{-1}$ & 41 & 1204 & 498 & 44 & 0.380 \% \\ 
    \hline
    $3.10^{-2}$ & 41 & 1372 & 498 & 44 & 0.380 \% \\ 
    \hline
    $1.10^{-2}$ & 41 & 1315 & 498 & 45 & 0.380 \% \\ 
    \hline
    $1.10^{-3}$ & 41 & 1531 & 498 & 45 & 0.380 \% \\ 
    \hline  

     \hline \hline
     \multicolumn{6}{|c|}{EXPERIMENT B} \\
     \hline 
     $k_\textrm{Res} $ & \textbf{CG} & \textbf{CG}  & \textbf{Newton} & \textbf{CPU} & \textbf{Solution} \\
     & \textbf{corrections} & \textbf{iterations}  & \textbf{iterations} & \textbf{time} & \textbf{error} \\ 
    \hline \hline
    $10^{0}$ &  31 & 827 & 384 & 36 & 0.894 \% \\ 
    \hline
    $10^{2}$ & 32 & 895 & 410 & 36 & 0.887 \% \\ 
    \hline
    $10^{4}$ & 32 & 896 & 453 & 39 & 0.887 \% \\ 
    \hline
    $10^{6}$ & 32 & 896 & 500 & 42 & 0.887 \% \\  
    \hline

  \end{tabular}
\end{center}
\caption{Influence of the parameters of the C-POD on the accuracy and efficiency of the corrective model order reduction}
\label{table:influenceCG}
 \end{table}

\subsection{Number of additional vectors in the reduced basis}

A new set of numerical experiments are performed in order to study the relevancy of the adaptive reduced model created during the successive time steps. The following set of parameters fixed parameters is chosen: $\critnew=10^{-1}$, $\critCG=\critnew$,  $k_\textrm{Res}=10^{3}$ and the value of the maximum number of solutions added to the reduced model, $n_\textrm{C,NewT,Max}$ is incremented. The results obtained are given in table (\ref{table:influence_kres}). The following general tendencies appear:
\begin{itemize}
\item The number of correction steps required to obtain a given accuracy drastically decreases as the size of the reduced model (number of additional reduced basis vectors) increases.
\item The global CPU time rapidly decreases when using an increasing, but small, number of additional reduced basis vectors. Though, when too high a value is given for $n_\textrm{C,NewT,Max}$, the costs of the assembly steps becomes dominant, and the numerical efficiency decreases.
\end{itemize}

\begin{table}[ht]
\begin{center}
  \begin{tabular}{ | c || c | c | c | c |c |  }
    \hline

     $n_{\textrm{C,NewT,Max}}$ & \textbf{CG} & \textbf{CG}  & \textbf{Newton} & \textbf{CPU} & \textbf{Solution} \\
     & \textbf{corrections} & \textbf{iterations}  & \textbf{iterations} & \textbf{time} & \textbf{error} \\ 
    \hline \hline
    0 &  74 & 2696 & 560 & 60 & 0.780 \% \\ 
    \hline
    1 & 45 & 1367 & 435 & 40 & 0.773 \% \\ 
    \hline
    3 & 32 & 896 & 429 & 38 & 0.887 \% \\ 
    \hline
    10 & 14 & 370 & 409 & 36 & 1.02 \% \\  
    \hline
    20 & 14 & 360 & 418 & 41 & 1.02 \% \\  
    \hline

  \end{tabular}
\end{center}
\caption{Influence of $n_{\textrm{C,NewT,Max}}$ on the efficiency of the corrective algorithm model order reduction}
\label{table:influence_kres}
 \end{table}

\subsection{Results}
\label{sec:First_results}

We can finally evaluate the suitability of the adaptative algoritm to the purpose for which it has been designed, namely the automatic correction of a reduced model.

The previously studied parameters are fixed to the following values: $\critCG=\critnew$,  $k_\textrm{Res}=10^{3}$ and $n_\textrm{C,NewT,Max}=3$. Several simulations are performed for different values of the control parameter $\critnew$ (interpreted as the error criterion on the reduced model). 
The results are reported in table (\ref{table:results_crack}). For comparison, the results obtained with the basic POD strategy and for a classical Newton algorithms on the complete system are also given.

\begin{table}[ht]
\begin{center}
  \begin{tabular}{ | c || c | c | c | c |c |  }
    \hline

     & \textbf{CG} & \textbf{CG}  & \textbf{Newton} & \textbf{CPU} & \textbf{Solution} \\
     & \textbf{corrections} & \textbf{iterations}  & \textbf{iterations} & \textbf{time} & \textbf{error} \\ 
    \hline \hline
    POD &  &  & 266 & 17 & 8.27 \% \\ 
    \hline
    $\critnew=8.10^{-1}$ & 4 & 46 & 274 & 21 & 4.11 \% \\ 
    \hline
    $\critnew=3.10^{-1}$ & 12 & 262 & 334 & 26 & 1.61 \% \\ 
    \hline
    $\critnew=1.10^{-1}$ & 32 & 896 & 429 & 38 & 0.887 \% \\ 
    \hline
    $\critnew=3.10^{-2}$ & 54 & 1601 & 556 & 52 & 0.263 \% \\ 
    \hline
    $\critnew=1.10^{-2}$ & 73 & 2107 & 705 & 67 & 0.157\% \\ 
    \hline
    $\critnew=1.10^{-3}$ & 132 & 4351 & 1539 & 144 & 0.00424 \% \\ 
   \hline
    Full Newton &  &  & 824 & 434 &  \\ 
    \hline

  \end{tabular}
\end{center}
\caption{Efficiency and accuracy of the corrective model order reduction}
\label{table:results_crack}
 \end{table}

\begin{figure}[htb]
       \centering
       \includegraphics[width=1 \linewidth]{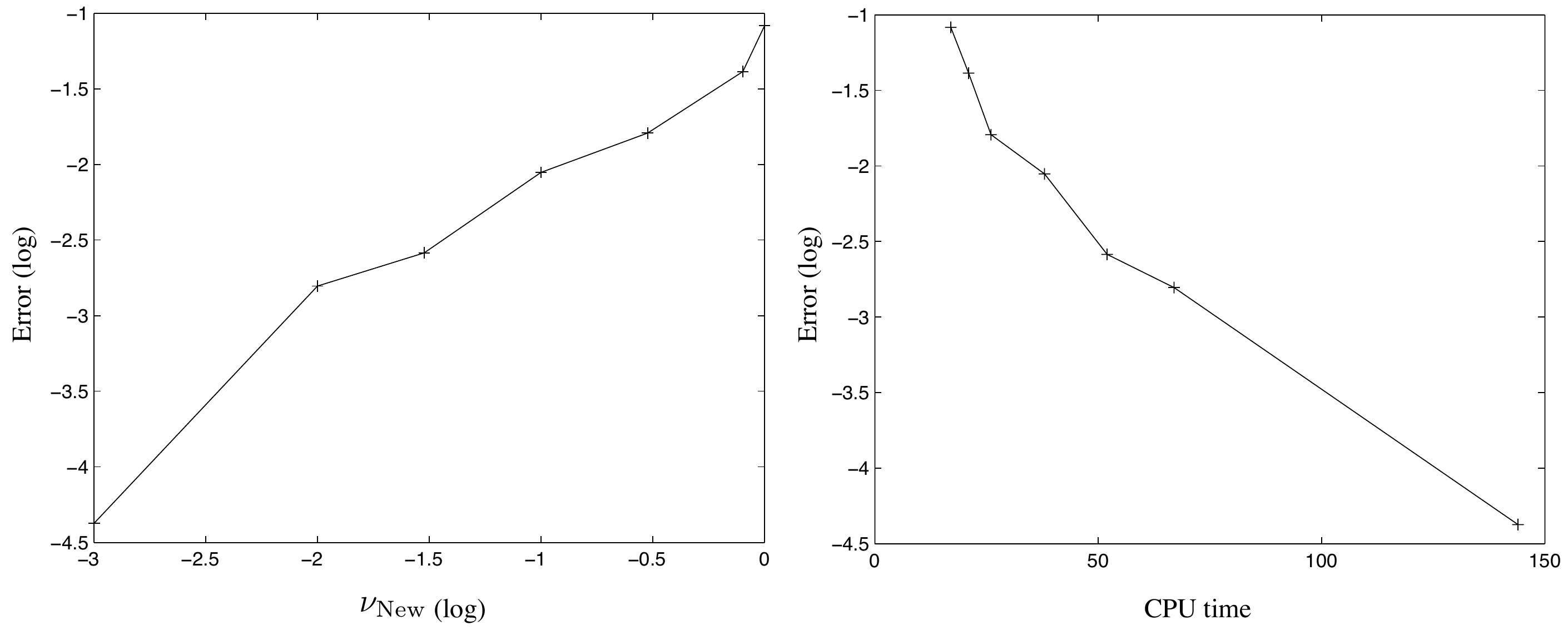}
       \caption{Solution errors obtained with decreasing values of $\critCG$, for $\critnew = 4.10^{-1}$ (left) and $\critCG = 2.10^{-1}$ (right). }
       \label{fig:Error_CPU_time}
\end{figure}

The first global observation is that the numbers of corrections and cumulated Newton iterations increases as $\critnew$ decreases. As a consequence, the CPU time increases with decreasing values of the control parameter. More importantly, the accuracy of the solution is controlled by $\critnew$. Figure (\ref{fig:Error_CPU_time}, left) shows that the relation between the solution error and the threshold value is almost linear (log-log scaling). Similarly, figure (\ref{fig:Error_CPU_time}, right) demonstrates that we obtain a linear relation between the accuracy of the solution and the CPU time required for the computation. Hence, the adaptive algorithm acts as a continuous corrective link between the basic POD strategy on the one hand, and a fine solution scheme for the full nonlinear problem on the other hand.

\begin{figure}[htb]
       \centering
       \includegraphics[width=1. \linewidth]{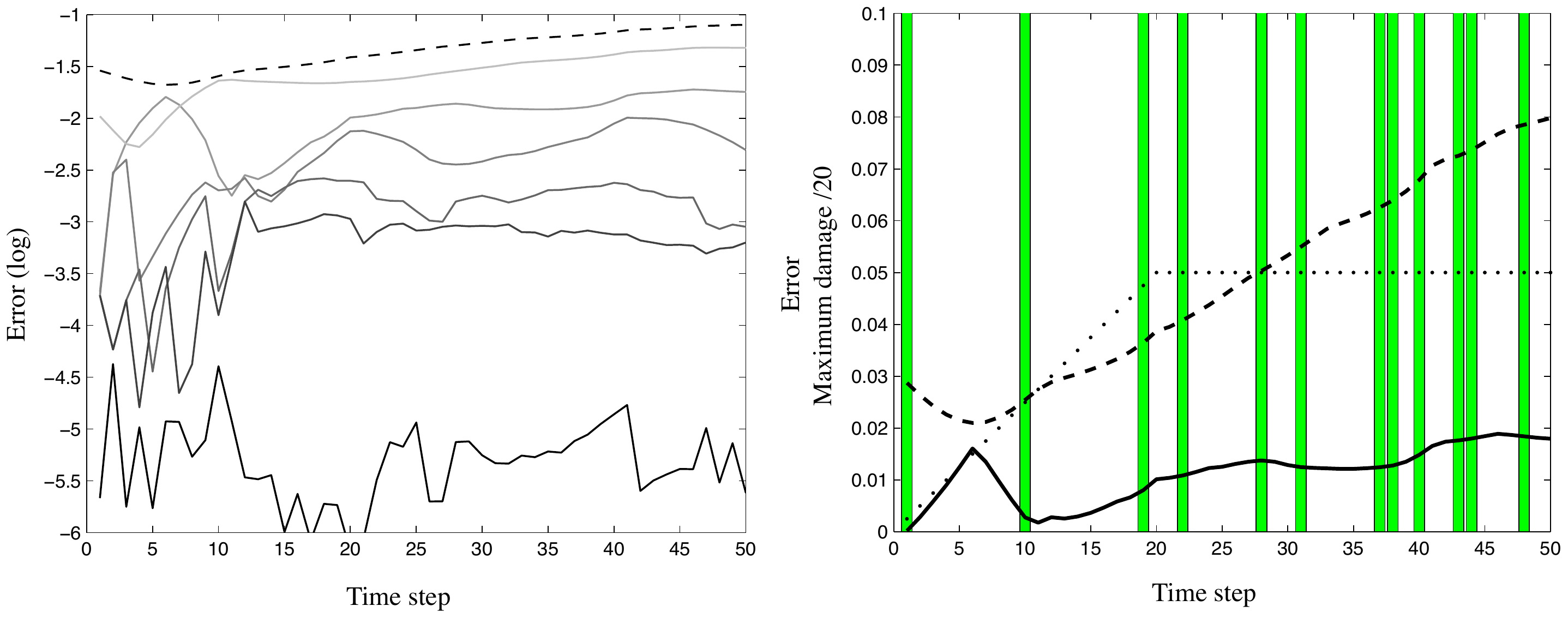}
       \caption{Error in solution for different values of the residual error of the initial problem $\critnew$ (left). Darker curves correspond to lower values of $\critnew$. The dashed curve is obtained by applying the basic POD and increases significantly with time. 
When using the corrective POD (C-POD), the error decreases in a monotonic manner with decreasing values of $\critnew$. On the right, the dotted line is the maximum damage over the structure. The dashed line is still obtained by applying the basic POD, is reasonably low in the initiation phase (no significant topological changes), and increases when damage propagates. The bar corresponds to the time steps being corrected when applying the C-POD ($\critnew=3.10^{-1}$). The frequency of correction increases when strong topological changes occur. The dark line is the resulting error in solution, and is kept to a low level.}
       \label{fig:Error_New_CG}
\end{figure}

The results detailed below help to understand how the successive corrections are performed during the simulation. Figure (\ref{fig:Error_New_CG}, left) presents the evolution of the solution error with time, for the different values of $\critnew$ used in this study. A ``zoom" in these results is plotted in figure (\ref{fig:Error_New_CG}, right), where we focus on the curve obtained by the application of the basic POD, and the one obtained by the adaptive algorithm, with $\critnew=3.10^{-1}$. The evolution of the maximum damage in the structure (linear because of the local arc-length control) is also reported. Each vertical bar point out the time steps where at least one correction of the reduced basis has been performed. One can see that a correction is necessary at the very beginning of the simulation, to take into account the specificity of the loading. The corrected reduced model constructed after the time step is sufficiently relevant to allow the simulation of the initiation of the damage in the structure, at the precision required by the predefined value of $\critnew$ (a single additional correction is performed at Time step 10). However, when the maximum damage exceeds 1 (Time step 20), the rate of corrective steps increases as the fast topological changes must be taken into account in the reduced model.

For illustration, and in the particular case detailed in the previous paragraph, the solution obtained at the end of the simulation is superimposed to the reference solution in figure (\ref{fig:Enriched_POD_reference}). The correction of the reduced basis is also plotted and shows an important deformation in the region where the damage localises. Elsewhere, the value of this vector is almost null, which means that, far away from the``crack", the displacement field is correctly represented by the initial reduced basis. In other terms, the correction is local, which obviously leads to the conclusion that multigrid or domain decomposition solvers could help to reduce the cost of the corrections. This issue will be discussed later on.

\subsection{Discussion}

We have shown in the previous section that the error-controlled POD-Krylov algorithm is a bridge between a full Newton/Krylov solver and the basic POD method. 
This algorithm can obviously be used to correct the reduced model of various engineering problems. Yet, in the case of mechanical models involving internal variables, 
the computation time required to assemble the reduced stiffness operator (evaluation of internal forces) is far too high to yield 
a computationally 
efficient strategy. Indeed, although the number of unknown is considerably reduced,
the evaluation of the internal variables and internal forces are done in each element of the structure.
In \cite{ryckelynck2008}, it is proposes to use the hyperreduction to reduce the cost of these operations. 
In the next section, we propose to apply the adaptive algorithm to the hyperreduction of damage models.


\section{Krylov corrections for the hyperreduction of damage models}

\label{sec:hyperred}

\subsection{Principle of the hyperreduction strategy}

The principle of the hyperreduction, which we will denote by H-POD, was first described in \cite{ryckelynck2005}, and applied to
mechanical problems involving internal variables in \cite{ryckelynck2008}. It drastically reduces the cost of evaluating
the internal forces required to compute the successive linearised operators and residuals in the Newton
algorithm for a problem reduced by the POD. The main idea is to consider only a few of the equations of the initial
problem to find the coefficients associated to each of the reduced basis vectors. If the controlled equations and the
reduced basis vectors are correctly chosen, the norm of the residual due to the equations which are not controlled will be small.
Note that we do not aim at questioning the validity of this strategy, but to use it as an efficient solver for
damage models, and test the efficiency of the adaptive reduced basis in such a framework

\subsection{Petrov-Galerkin formulation of the nonlinear system of equations}

The starting point of the strategy is the full system \eqref{eq:discretized_equilibrium}, 
in which the approximation of the displacement as a linear combination of reduced basis vectors has been introduced:
\begin{equation}
\label{eq:hyperreduced_intro}
\fint \left( \dispdiscr_{| t_n}+\basereduc \, \coeffreduc  \right) + \fext  = \zerodiscr
\end{equation}
This system of equations is obviously overconstrained. A solution to this system is obtained in the classical model order reduction
by using the Galerkin orthogonality condition with respect to the reduced basis. 
But one can choose any orthogonality conditions leading to a well-posed problem. A``natural" choice can be to satisfy
only a very few number of equations $n_h$ of \eqref{eq:hyperreduced_intro} (called controlled equations), with $n_h \geq n_\textrm{C}$
\begin{equation}
\label{eq:hyperreduced_intro2}
\M{\mathbf{E}}^T \left( \fint \left(  \dispdiscr_{| t_n}+\basereduc \, \coeffreduc  \right) + \fext \right) = \zerodiscr
\end{equation}
$\M{\mathbf{E}}$ is a Boolean operator of $n_u$ (number of nodal unknown) rows and $n_h$ columns. For each column of $\M{\mathbf{E}}$, the only 
non-zero value correspond to the $i^{th}$ controlled equation, for $i \in \llbracket 1 \ n_h \rrbracket$, and is set to 1.
 
In the H-POD, problem \eqref{eq:hyperreduced_intro2} is required to satisfy the Galerkin
orthogonality condition with respect to the reduced snapshot, which finally yields the following system:
\begin{equation}
\label{eq:hyperreduced}
\M{\mathbf{C}}^T \M{\mathbf{E}} \, \M{\mathbf{E}}^T \left( \fint \left(  \dispdiscr_{| t_n}+\basereduc \, \coeffreduc  \right) \right) + \M{\mathbf{C}}^T \M{\mathbf{E}} \M{\mathbf{E}}^T \, \fext = \zerodiscr 
\end{equation}

\subsubsection{Nonlinear solver}

The Newton Procedure described in section \ref{sec:sol_proc_red} can be applied to the hyperreduced problem \eqref{eq:hyperreduced}. 
At iteration $i$ of the Newton algorithm the following linear prediction is performed,
\begin{equation}
\optanghyperred^i \, \deltacoeffreduc^{i+1} = - \resdiscrhyperred^{i}
\end{equation}
where $\resdiscrhyperred^i = \basereduc^T \M{\mathbf{E}} \, \M{\mathbf{E}}^T  (\fext + \fint( \dispdiscr_{| t_n}+\basereduc \, \coeffreduc^i))$. The correction stage reads: 
\begin{equation}
\begin{array}{c}
\displaystyle \resdiscrhyperred^{i+1} = \basereduc^T  \M{\mathbf{E}} \, \M{\mathbf{E}}^T \fext + \basereduc^T  \, \M{\mathbf{E}} \, \M{\mathbf{E}}^T \fint( \dispdiscr_{| t_n}+\basereduc \, \coeffreduc^{i+1}) 
\\ 
\displaystyle {\optanghyperred^{i+1}} = \basereduc^T \, \left. \frac{\partial \, \M{\mathbf{E}} \, \M{\mathbf{E}}^T  \fint(  \dispdiscr_{| t_n}+\basereduc \, \coeffreduc )}{\partial \coeffreduc} \right|_{ \coeffreduc = \coeffreduc^{i+1}}
\end{array}
\end{equation}
The resulting linearized operator ${\optanghyperred^{i}}$ is square, fully populated, non-symmetric and of very small size $n_h \times n_h$. It is solved by a direct solver (LU factorization).

Note that $\M{\mathbf{E}} \, \M{\mathbf{E}}^T$ is not explicitly computed. It simply extracts the internal and external forces
of the controlled equations used to calculate the residuals and linearised operators. The integration of the internal
and external forces needs only be done on a reduced integration domain $\Omega_h$, made of all the elements connected to
a controlled node (see figure (\ref{fig:RID})).

The same developments can be done for the derivation of the arc-length procedure, which is used in the examples of this section.

\subsubsection{Reduced integration domain}

The choice of the controlled equations is the keypoint of the hyperreduction strategy. In \cite{ryckelynck2005,ryckelynck2008}, it is shown that in order all the reduced basis vectors to be ``observable'', one should choose to control the equations associated
to the nodes whose connected elements are subjected to the largest strain energy from the reduced basis vectors. Hence, it is recommended to perform a loop on the nodes and compute the average of the strain energy due to each 
individual reduced basis vector in the connected elements. Then, for each reduced basis vector, 
one chould choose a given, and small, number of nodes with the maximum average strain energy observed by the connected elements. Finally, the 
controlled equations should correspond to all the degrees of freedom of these particular nodes, and the connected elements define the reduced integration domain.

\subsection{Modified Error-controlled algorithm}


When solving the hyperreduced system by our adaptive algorithm (the resulting strategy will be called corrective hyperreduction, and denoted by CH-POD), certain operations become cost-inefficient:
\begin{itemize}
 \item The calculation of the residual $\resdiscr$.
 \item The assembly of the global stiffness, required to enrich the reduced basis.
 \item The conjugate gradient itself if not correctly preconditioned.
\end{itemize}
These issues are discussed next.

\subsubsection{Calculation of the damage variables and residual of the initial problem}

In the H-POD method, the calculation of the internal variables needs only be done on the reduced integration domain $\Omega_h$. In \cite{ryckelynck2008},
it is proposed to interpolate these internal variables over $\Omega \backslash \Omega_h$ thanks to pre-calculated snapshot functions. In our case, this particular
feature is not of interest, for:
\begin{itemize}
 \item The residual of the global system needs to be controlled.
 \item The residual of the global system needs to be calculated when required to perform an enhanced prediction.
\end{itemize}

Therefore, the internal variables are computed systematically
when the relative norm of the residual of the reduced problem reaches a treshold value. 
The norm of the full residual can thus be computed, and if its value is too high, 
an enhanced linear prediction (e.e.: a correction of the reduced basis by the Krylov solver) is performed. Note that this calculation is very expensive compared to the cost of the Newton iterations performed on the hyperreduced system, 
and must be limited to a minimum number.

Let us recall that in the former algorithm given in section \ref{sec:error_controlled}, an enhanced linear prediction could be performed if
$\displaystyle \frac{ \lVert \resdiscrred \rVert_2 }{ \lVert \M{\mathbf{C}}^T \fext \rVert_2} < k_{\textrm{Res}} \displaystyle \frac{ \lVert \resdiscr \rVert_2  }{ \lVert \fext \rVert_2 } $.
It requires the computation of the full residual at any iteration of the Newton solver, which is, in the hyperreduction strategy, computationally inefficient.

Hence, for this particular application, this comparison of the two norms is bypassed, and an enrichment of the reduced basis will not be made unless the relative norm of the
reduced problem is lower than $\critnewred$ (convergence criterion on the reduced nonlinear problem, set to a very low value), thus limiting the number of Newton iterations at which the internal damage variables must be calculated in $\Omega \backslash \Omega_h$. An another modification permits to suppress a large quantity of the numerical costs. The norm of the full residual is controlled at the first time step, and, if no correction is needed, no control is done during the following ${n_{C,T}}^{th} time step$ ($n_{C,T}=2$ in the following test cases). The same procedure is repeated after each controlled time step.

\subsubsection{Global stiffnesses operators}

Obviously, the assembly of the global stiffness is the weak point of this error-controlled hyperreduction strategy. Yet, as the conjugate gradient algorithm is only used to enrich the reduced basis, one does not need to use the current stiffness matrix. 
The ``naive'' technique to suppress the costs associated with this operation would then be to use a stiffness matrix assembled during the computation of the snapshot.
Though, as the damage in the structure increases, the relevance of the computed additional reduced basis vectors can be questioned, leading to a slower convergence of the adaptive
strategy, which will be illustrated later on.

We propose an intermediate technique called the``patch assembly" (PA). Before an enhanced prediction stage, the residual of the full system is known. We can reasonably
assume that the part of the domain $\Omega_\textrm{PA}$ of highest local residuals correspond to high levels of structural changes. Therefore, we get an approximation of the current stiffness operator
by removing the previous elementary stiffnesses corresponding to the elements in $\Omega_\textrm{PA}$
from the last stiffness approximation, and replacing them by their updated values.

This procedure is performed in three steps:
\begin{enumerate}
 \item Identify the nodes corresponding to the highest values of the residual.
 \item For each of this nodes, identify the connected elements, whose union defines $\Omega_\textrm{PA}$.
 \item In the existing approximation fo the stiffness operator $\optanghat^{i}$, remove the elementary matrices corresponding to the elements in $\Omega_\textrm{PA}$:
 \begin{equation}
  \optangtilde = \optanghat^{i} - \sum_{E/ \Omega_E \in \Omega_\textrm{PA}} \M{\mathbf{A}}_E^T \, \optangE(Z_\textrm{Old}) \, \M{\mathbf{A}}_E
 \end{equation}
  where $Z_\textrm{Old}$ represent the state variables and material parameters used to assemble $\optang^{i}$, $(\M{\mathbf{A}}_E)_{E \in \llbracket 1 \, n_E\rrbracket}$ are Boolean assembly operators, which extract the rows of the global stiffness corresponding to the degrees of freedom of Element E, and
$\optangE(Z_\textrm{Old})$ the elementary stiffness operator.
 \item Calculate the elementary matrices corresponding to the elements in $\Omega_\textrm{PA}$ and assemble them in $\optangtilde$:
 \begin{equation}
  \optanghat^{i+1} = \optangtilde + \sum_{E/ \Omega_E \in \Omega_\textrm{PA}} \M{\mathbf{A}}_E^T \, \optangE(Z^{i}) \, \M{\mathbf{A}}_E
 \end{equation}
\end{enumerate}

The performance of this simple technique will be demonstrated later on.

\subsubsection{Preconditionning}

As the size of the initial problem increases, the preconditioning of the conjugate gradient used to enrich the reduced basis becomes an issue. 
In order to be consistent with the projection framework that we have introduced, we make use of the Selective Reuse of Krylov Subspace (SRKS), proposed in \cite{gosselet2003}.
Briefly, the preconditioner itself is unchanged (diagonal matrix whose entries are the inverse of the diagonal elements of the stiffness operator), but additional reduced basis vectors are added to the augmentation space:
\begin{equation}
\widetilde{\basereduc} = 
\left( \begin{array}{cc}
\basereduc & \basereduc_\textrm{SRKS} 
\end{array} \right)
\end{equation}
The columns of $\basereduc_\textrm{SRKS}$ are made of eigenvectors of the successive linearised operators projected on $\textrm{Im}(\widetilde{\basereduc}^{\bot_{\optang}})$ during the computation of the snapshot.
These eigenvectors are the converged Ritz vectors obtained when solving the linear systems by a conjugate gradient (eigenvectors of the complete linearised operator associated with the highest eigenvalues).

This feature permits to rapidly access the super-convergence of the conjugate gradient.

\subsubsection{Hyperreduced arc-length strategy}

The search for the largest increment of displacement in equation \eqref{eq:control_eq} is only done on the reduced integration domain. 
A sufficient number of point
corresponding to the highest damage rate is set as controlled nodes at the end of each time step. Therefore, one can expect that the control equations is 
reasonably satisfied.

\subsection{Efficiency for parametric simulations}

\subsubsection{Presentation of the test cases}

\begin{figure}[htb]
       \centering
       \includegraphics[width=1.0 \linewidth]{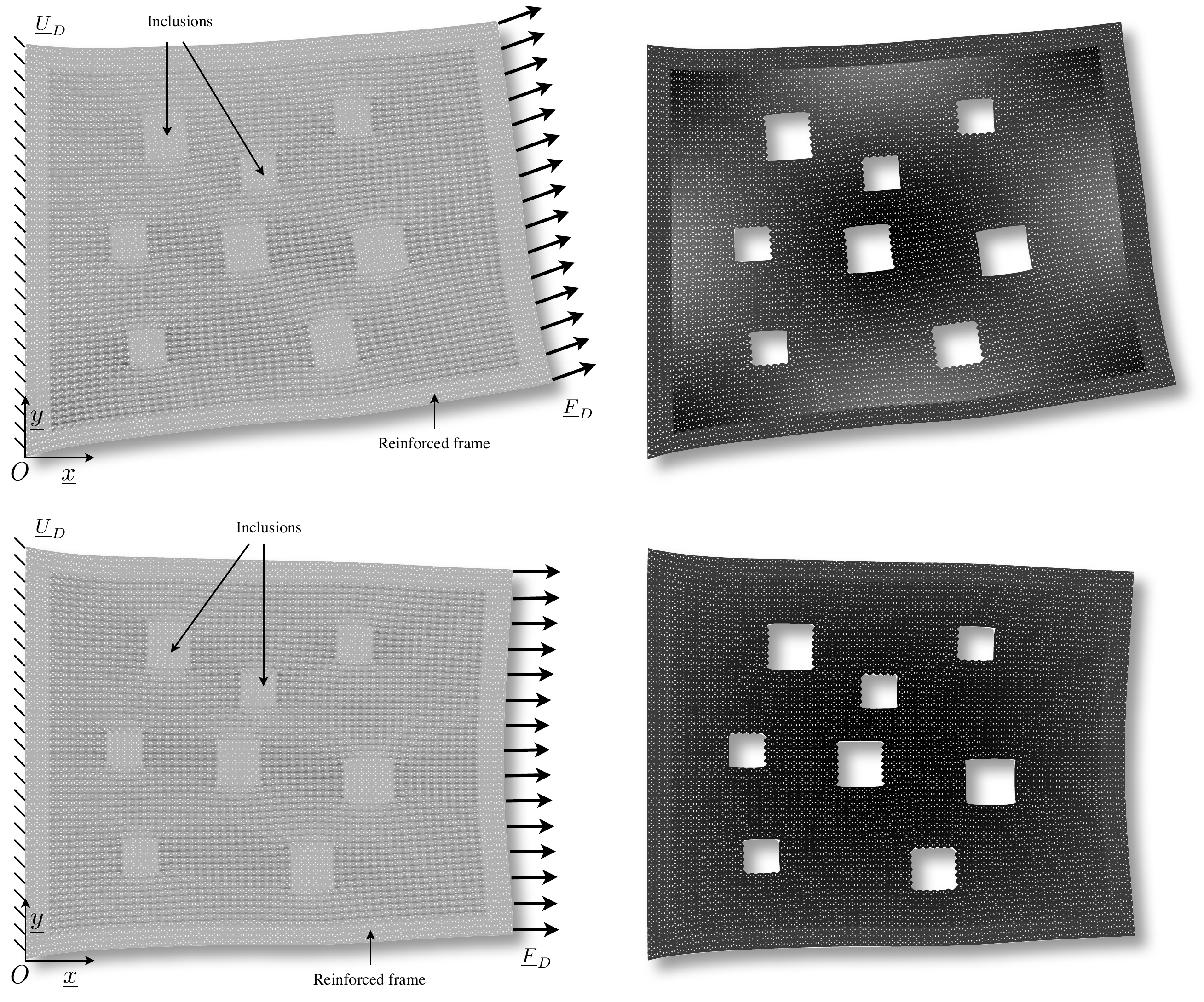}
       \caption{Reference problem and associated solutions obtained for different material parameters and direction of the prescribed load (left), where darker colours correspond to higher value of damage, and Young's modulus cosinusoidal distribution in these two cases (right), where darker colours correspond to higher values of the stiffness parameter.}
       \label{fig:Reference_large}
\end{figure}

The test case studied here is a square damageable lattice (the constitutive modelling is the one given in section \ref{sec:lattice}) represented in figures (\ref{fig:Reference_large}). The lattice structure is made of 14,520 bars 
connected by 3721 nodes. Hence the number of degree of freedom involved in this problem is $3721 \times 2 = 7442$.
A non-zero  homogeneous traction force such that $\Fd = F_x \,\V{x} + F_y \,\V{y}$ is applied on the right edge of the
structure, while zero-displacement are prescribed on its right edge. We define the angle:
\begin{equation}
 \phi = \arctan \left( \frac{F_y}{F_x} \right)
\end{equation}

The load is applied
through a reinforced frame occupying domain $\Omega_\textrm{F}$. The Young's Modulus of the bars belonging to the frame is equal to one, while their damage parameter $\alpha$ is set to $0.1$. 
Hard inclusions, occupying domain $\Omega_\textrm{I}$ are located in domain $\Omega \backslash \Omega_\textrm{F}$ (see their position and shape on figure (\ref{fig:Reference_large})). The Young's Modulus of the bars such that 
$\Omega_b \in \Omega_\textrm{I}$ is set to $10$ while $\alpha=0.01$. Due to these low values of $\alpha$ in $\Omega_\textrm{F} \cup \Omega_\textrm{I}$, the damage will localize in the complementary
domain $\Omega_\textrm{C} = \Omega \backslash (\Omega_\textrm{F} \cup \Omega_\textrm{I})$, where $\alpha$ is set to $1$. The young modulus of the bars in this domain is defined by the following law:
\begin{equation}
E|_{\Omega_b \in \Omega_\textrm{C}} = E_\textrm{C} \left( 1 + \epsilon_\textrm{C} \left( \sin \left( \omega_\textrm{C} \left( (X_b-\bar{X})+(Y_b-\bar{Y}) \right) \right) + \sin \left( \omega_\textrm{C} \left( (X_b-\bar{X})-(Y_b-\bar{Y}) \right) \right)  \right) \right)
\end{equation}
where the position of the barycentre $G_b$ of a bar occupying domain $\Omega_\textrm{b}$ is 
\begin{equation}
\V{OG_b} = X_b \, \V{x} + Y_b \, \V{y}
\end{equation}
and the position of barycentre $G$ of the structure is defined by:
\begin{equation}
\V{OG} = \bar{X} \, \V{x} + \bar{Y} \, \V{y}
\end{equation}

The purpose of our study is to obtain the dissipated energy in the structure, for a value of the load that is close to the instability point (hence the arc-length control is not required), as a function of the two following parameters:
\begin{itemize}
\item the angle $\phi$, which controls the direction of the force.
\item the angular frequency $\omega_\textrm{C}$, which controls the spatial distribution of the Young's modulus of the bars belonging to $\Omega_\textrm{C}$.
\end{itemize}
The construction of such functions is very classical is stochastic analysis. It is often associated with the Monte-Carlo method to determine the random distribution of a quantity of interest (in our case the dissipated energy) as a function of the stochastic input parameters (here $\phi$ and $\omega_\textrm{C}$). If an analytical derivation of this function is not possible, the classical procedure consists in performing a few deterministic simulations, for different values of the input parameters, and interpolate the value of the quantity of interest between these points. 

\begin{figure}[htb]
       \centering
       \includegraphics[width=0.5 \linewidth]{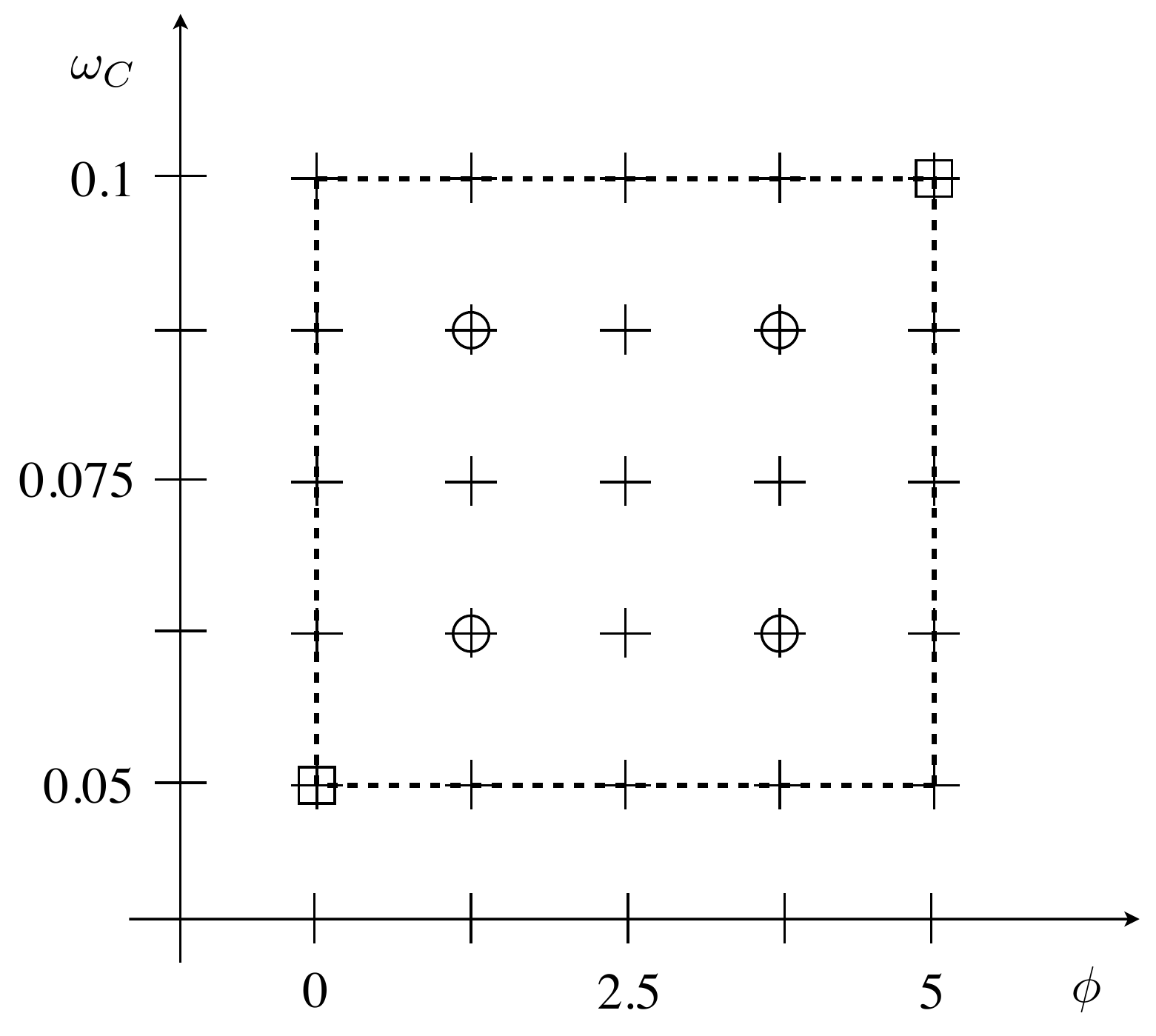}
       \caption{Estimation of the response surface. The surface relates the dissipated energy to two structural parameters $\phi$ and $\omega_\textrm{C}$ over a rectangular domain. The values of this function at 25 nodes (crosses) are computed by performing nonlinear simulations. The other values are interpolated using a piecewise linear approximation.}
       \label{fig:RSE}
\end{figure}

Let us recall that the dissipated energy of the chosen damage model is given by: 
\begin{equation}
E_{\textrm{Dissi}} = \int_{\Omega} \left( \int_{t=0}^{T} \frac{N}{S_b}  \dot{\epsilon} \, dt - \frac{1}{2} E(1-d|_{t=T}) \epsilon^2 \right) \, d \Omega = \int_{\Omega} \int_{t=0}^{T}  \frac{Y}{S_b}  \dot{d}   \, dt \, d \Omega
\end{equation}
The following values are chosen for the constant parameters: $E_\textrm{C}=1$, $\epsilon_\textrm{C}=0.2$. Particular distributions of the Young's modulus are represented on figure (\ref{fig:Reference_large}, right).

\begin{figure}[htb]
       \centering
       \includegraphics[width=1. \linewidth]{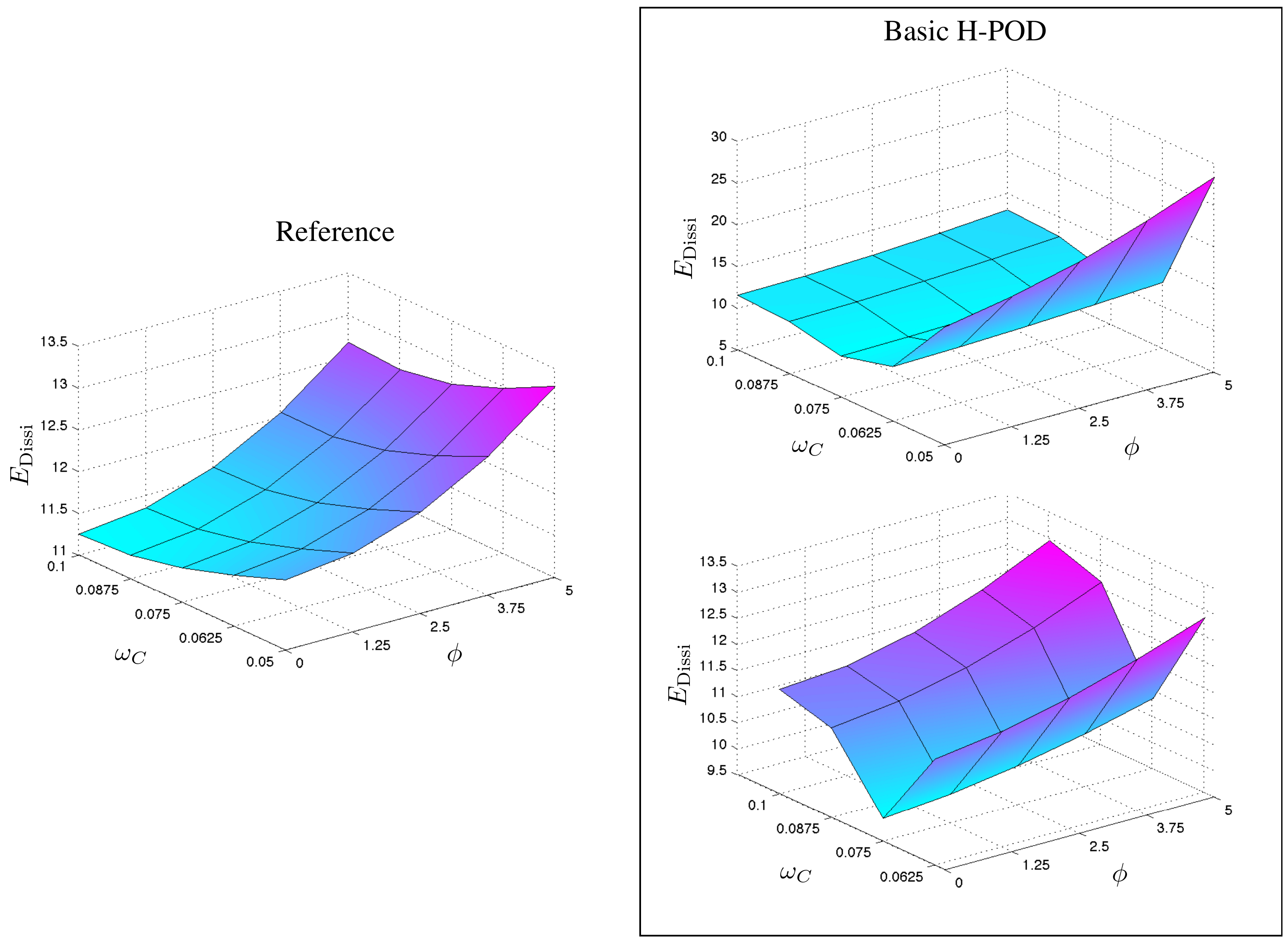}
       \caption{Response surface obtained in the reference case, and when using the H-POD. The error in the dissipated obtained when not correcting the reduced model is very high, due to the large distance between the solution which is searched for and the \textit{a priori} computed snapshot.}
       \label{fig:surf_Ref_POD}
\end{figure}

We will consider that the input parameters can vary within the following range:
\begin{equation}
\left\{ \begin{array}{l}
\omega_\textrm{C} \in [\omega_\textrm{C,1} , \omega_\textrm{C,2}] = [0.05 , 0.1] \\
\phi \in [\phi_{1} , \phi_{2}] = [0 , 5]
\end{array} \right.
\end{equation}
25 simulations are performed to estimate the response surface (see figure \ref{fig:RSE}):
\begin{equation}
\left\{ \begin{array}{l}
\omega_\textrm{C} \in \{ 0.05 , 0.625 , 0.75 , 0.875 , 0.1 \}  \\
\phi \in \{ 0 , 1.25 , 2.5 , 3.75 , 5 \}
\end{array} \right.
\end{equation}
The response surface (linear interpolation) obtained when using a direct Newton solver over 10 time increments is given in figure (\ref{fig:surf_Ref_POD}, left). The displacement solutions, damage maps and Young's modulus distributions corresponding to the two limit points boxed in figure (\ref{fig:RSE}) are represented in figure (\ref{fig:Reference_large}).
Our goal is to limit the costs of these nearby simulations by using POD-based reduced order model for these 25 successive solutions, and in particular to test the efficiency of the CH-POD.

\subsubsection{Parameters of the hyperreduced model}

In order to build a snapshot, we solve successively four problems circled in figure \ref{fig:RSE}. Each of these simulations is performed in 10 time increments. The final snapshot, made of 40 vectors, is reduced to 8 reduced basis vectors by a singular value decomposition (the ratio of the first singular value of 
$\M{\mathbf{S}}$ over the eighth is $1.9\,10^{-8}$, which shows that the snapshot is correctly represented in $\basereduc$).

\begin{figure}[htb]
       \centering
       \includegraphics[width=0.45 \linewidth]{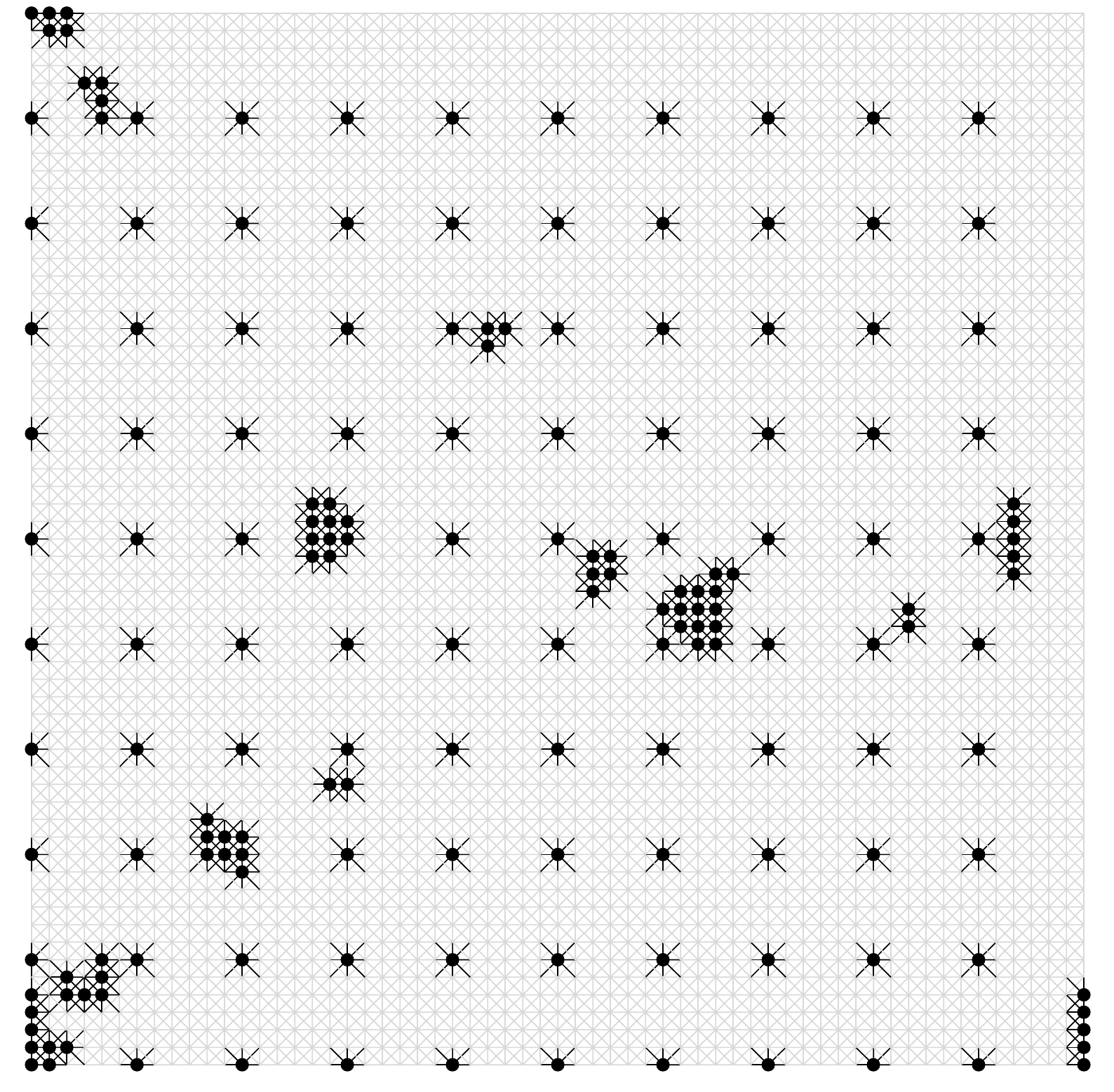}
       \caption{Reduced integration domain (dark bars) obtained at the end the first simulation performed to estimate the response surface. During the Newton process, the integration of the internal forces need be done only on this domain, which drastically reduces the computation costs.}
       \label{fig:RID}
\end{figure}

We construct the reduced integration domain by choosing a list of controlled nodes (2 controlled balance equations per node).
\begin{enumerate}
 \item The domain is divided into 100 square subdomains. For each of these subdomains, an arbitrarily chosen node is set as controlled point.
 \item 5 nodes are arbitrarily (first ten in the global nodal numbering) chosen for each of the prescribed boundary conditions (left and right edge of the lattice).
 \item For each vector of the reduced basis, 5 nodes whose connected element have the highest strain energy are set as controlled nodes.
 \item 20 additional nodes with the highest damage rate are added to the list, hence allowing to precisely integrate the constitutive law in this particular elements.
\end{enumerate}
An example of RID obtained during the CH-POD process is given in figure (\ref{fig:RID}).

\subsubsection{Results}

The following solvers are to estimate the response surface:
\begin{itemize}
 \item Newton algorithm on the complete system, the linear prediction being solved by a direct solver, and $\critnew = 10^{-6}$.
 \item ``basic'' hyperreduction (H-POD) with the snapshot and controlled points defined previously.
 \item corrective hyperreduction (CH-POD), where $\critnewred = 10^{-6}$ (threshold value on the relative norm of the residual of the reduced problem), $n_\textrm{C,NewT,Max}=3$ (number of maximum additional vectors added in the reduced basis during the Newton iterations) and $\critnew$ successively set to (a) $10^{-1}$ ,(b) $3.10^{-2}$ and (c) $10^{-2}$ (error criterion on the reduced model).
\end{itemize}


For each of these solvers we report the total CPU time required to construct the surface, and the maximum error in the dissipated energy over the 25 simulations (see table (\ref{table:dissipation})). This latter error is defined as follows:
\begin{equation}
\textrm{error} = \frac{\textrm{E}_{\textrm{Dissi} |_\textrm{Reference}} - \textrm{E}_{\textrm{Dissi} |_\textrm{Current experiment}}} {\textrm{E}_{\textrm{Dissi} |_\textrm{Reference}}}
\end{equation}

\begin{table}[htb]
\begin{center}
  \begin{tabular}{ | c || c | c |  }
    \hline

     & \textbf{Maximum error} &  \textbf{CPU time}   \\ 
     & \textbf{in dissipation} &  \textbf{(s)}   \\ 
    \hline \hline 
     \textbf{Full Newton (reference)} &    & 3227   \\ 
    \hline
    \textbf{H-POD} &  113  \% & 618   \\ 
    \hline
    \textbf{CH-POD} (a) &  5.06 \% &  730   \\ 
    \hline
    \textbf{CH-POD} (b) &  1.98 \% &  890   \\ 
    \hline
    \textbf{CH-POD} (c) &  0.82 \% &  1075   \\ 
    \hline

  \end{tabular}
\end{center}
\caption{Error and CPU time obtained when estimating the response surface with the H-POD and CH-POD compared to a reference response surface obtained by a classical Newton algorithm on the complete system of equations.}
\label{table:dissipation}
 \end{table}

\begin{figure}[p]
       \centering
       \includegraphics[width=1. \linewidth]{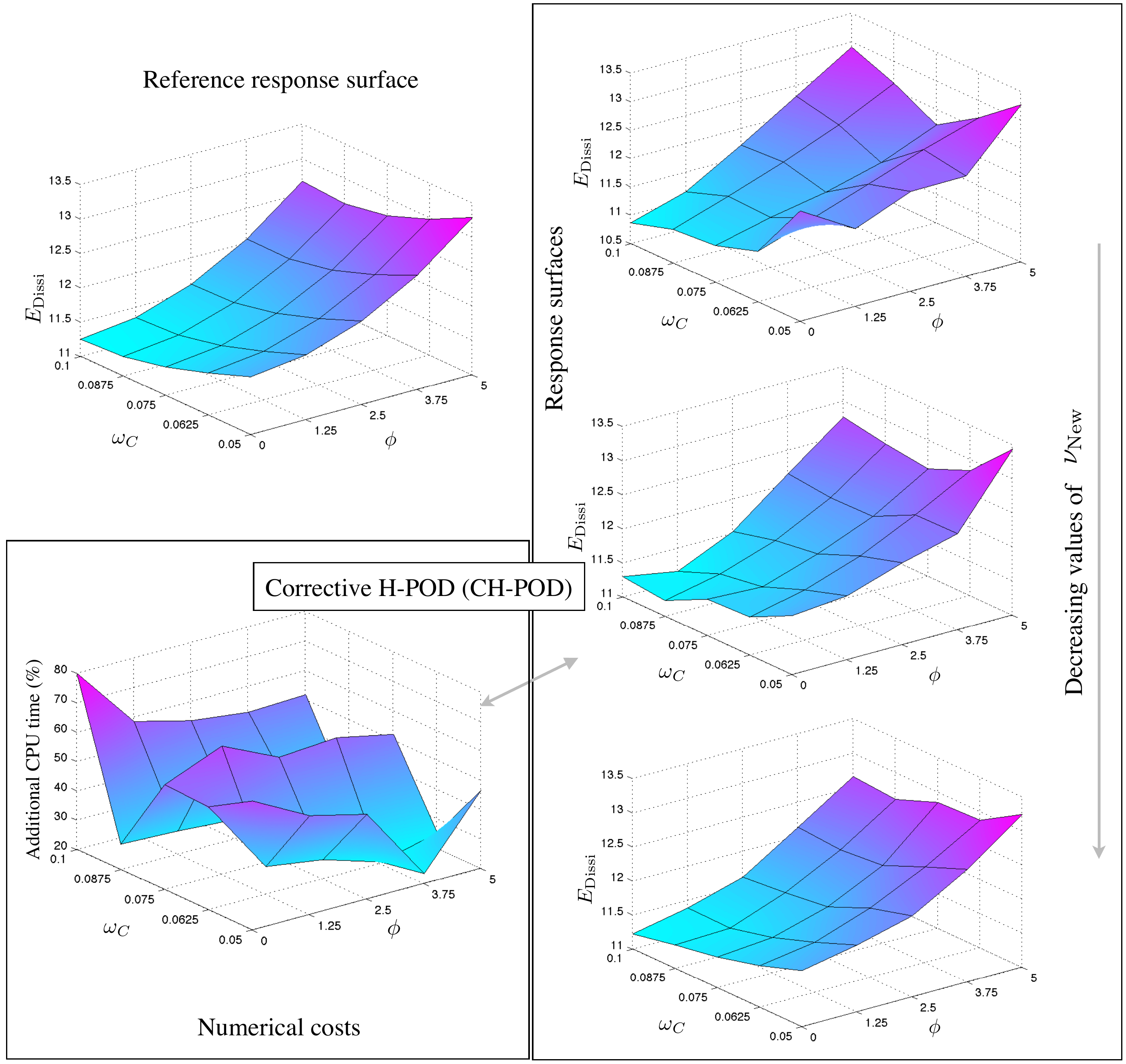}
       \caption{Response surface obtained in the reference case, and when using the CH-POD for various values of the reduced model error criterion, and CPU time for each of the 25 simulations (additional CPU times with respect to the ones obtained by applying hyperreduction). The error in the dissipated energy monotonically drops when the criterion decreases. The additional CPU time is concentrated in the simulations whose solution is far away from the snapshot.}
       \label{fig:surf_CH-POD}
\end{figure}

The results given by the hyperreduction are not satisfying. The maximum error obtained exceeds 100\%, which can be explained by observing the results presented in figure (\ref{fig:surf_Ref_POD}). One can see that the values of dissipated energy obtained for values of $\omega_\textrm{C}$ which have not not been taken into account in the snapshot are wrongly predicted, and especially for $\omega_\textrm{C}=0.05$. In this latter case, the structure is a lot weaker (greater quantity of bars with Young's modulus lower than $E_\textrm{C}$, as appears in figure (\ref{fig:Reference_large})) than in the cases computed to create the snapshot. As a consequence, the dissipated energy is overestimated.

The application of the CH-POD permits to retrieve a correctly estimated response surface for small additional costs. For instance, in experiment (a) ($\critnew=10^{-1}$), the maximum error drops to 5\% for 18\% of additional CPU time compared to the basic reduced order simulation. The response surface obtained is given in figure (\ref{fig:surf_CH-POD}).

Decreasing values of the reduced model error criterion ($\critnew$) results in a decreasing  value of the error, while slightly increasing the additional CPU costs (see figure (\ref{fig:surf_CH-POD}) and table (\ref{table:dissipation})). In experiment (c), the difference between the response surface obtained and the reference one can hardly be notice, and the evolution of the quantity of interest with respect to the input parameters is correctly taken into account. Yet, this simulation is less than two times as costly as the one made using the H-POD.

The CPU time, corresponding to each of the 25 simulations, in case (b), is also reported in figure (\ref{fig:surf_CH-POD}). The additional costs are due to the solution which is searched for to be far away, in some sense, from the snapshot. As explained previously, the Young's modulus distribution leads to the snapshot being irrelevant, which explains why the costs of the corrections in the CH-POD increases. Conversely, the variation of the angle, for $\phi>0$ is correctly taken into account in the uncorrected model order reduction, and the CH-POD generates very few extra costs in these cases. In the particular case where $\phi=0$ the solution is not correctly represented by a linear combination of the initial reduced basis vectors. Indeed, a strictly positive angle in the direction of the load leads to the localization of damage in one of the corner of the square structure (see figure \ref{fig:Reference_large}), while the structure is much stiffer for $\phi=0$. The CH-POD automatically corrects the resulting reduced model.

\subsubsection{Brief validation of the Patch-assembly}

In order the structural changes to be significant in the structure, Experiment A is performed with a higher value of the norm of the final loading (same value of the arc-length, but greater number of time increments). The evolution of the damage leads to the apparition of cracks in the vicinity of two of the hard inclusions.

We use three different approximation of the global stiffness operator in the ECH-POD strategy. The threshold value $\critnew$ is set to $10^{-1}$. In the first case, the stiffness is updated at the beginning of each time increment, which, as described previously, leads to an inefficiency due to the costs of the global assembly. In the second case, the initial stiffness is used to perform the enhanced linear prediction. The CPU time slightly decreases but the number of conjugate gradient calls required to obtain a relevant reduced basis increases significantly. 

\begin{figure}[htb]
       \centering
       \includegraphics[width=1. \linewidth]{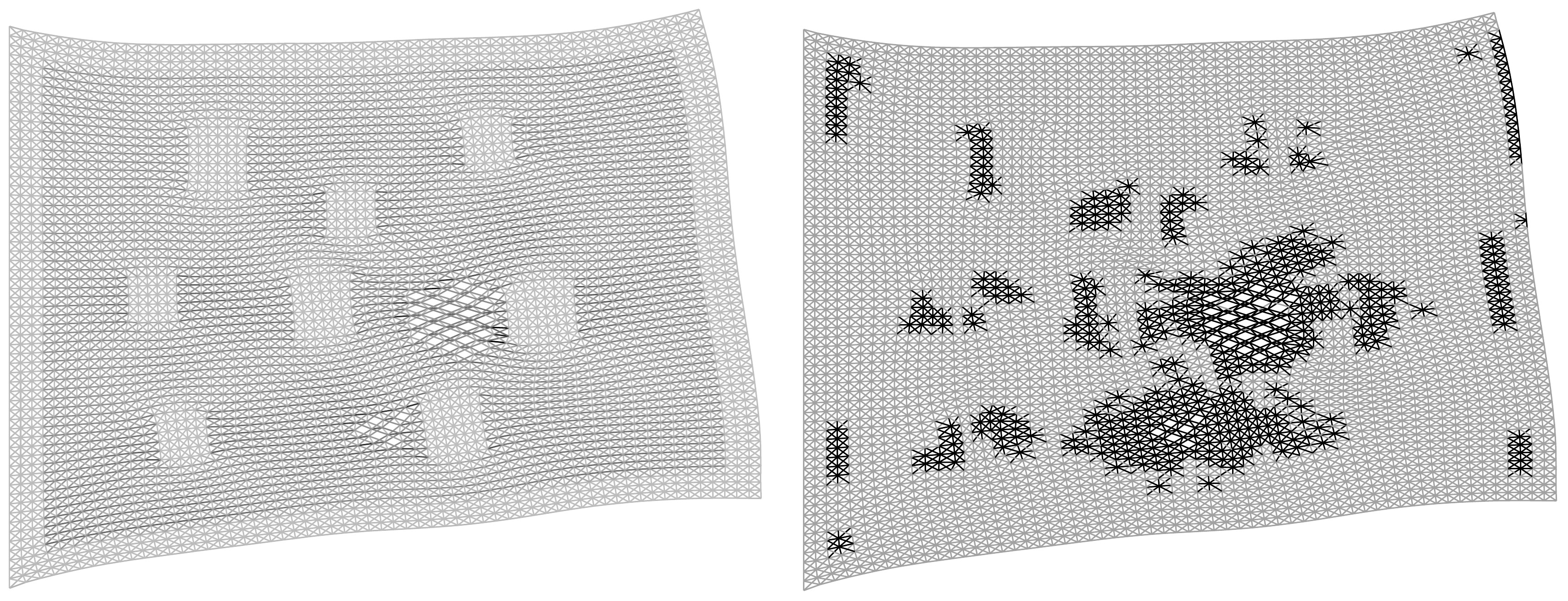}
       \caption{Solution obtained using the arc-length procedure after the limit point (left). On the right, the black bars are connected to the nodes corresponding to the highest values of the residual of the complete nonlinear system. They are mainly located in the zone of high topological changes.}
       \label{fig:Patch}
\end{figure}

Finally, the Patch-assembly technique is used to yield successive approximation of the global stiffness matrix. The number of elements in $\Omega_{PA}$ is 2440 (in comparison with the total number of bars: 14250), which gives a decrease of factor 6 of the costs of the global assembly. Figure (\ref{fig:Patch}) shows Domain $\Omega_{PA}$, which is indeed mainly located in the zones of high structural changes (the two cracked regions). The resulting CPU time is further decreased compared to the second experiment, as shown in table \ref{table:results}. Note that the number of conjugate gradient calls does not increase compared to the one obtain when updating systematically the stiffness operator, which shows that the approximation of this matrix is satisfying.

\begin{table}[htb]
\begin{center}
  \begin{tabular}{ | c || c | c | c | c |  }
    \hline
     & \textbf{CG} & \textbf{CG}  & \textbf{Newton} & \textbf{CPU}  \\
     & \textbf{corrections} & \textbf{iterations}  & \textbf{iterations} & \textbf{time}  \\ 
    \hline \hline
    Updated stiffness & 23  & 424  & 414 & 421  \\ 
    \hline
    Initial stiffness & 33 & 500 & 412 & 407  \\ 
    \hline
    Patch-assembly  & 22 & 440 & 447 & 335 \\
    \hline
  \end{tabular}
\end{center}
\caption{Efficiency of the Patch-assembly}
\label{table:results}
 \end{table}

\subsection{Discussion}

\subsubsection{For further efficiency}

A few important issues can severely limit the efficiency of the CH-POD derived in this section.
\begin{itemize}
\item The efficiency of the H-POD itself can be put into question for softening behaviour. Indeed, its numerical complexity is directly dependent on the size of the reduced integration domain. We have observed that choosing this domain in order to ensure the stability of the algorithm was not a simple issue, and chosen to use a large number of controlled nodes. By a better understanding of this behaviour, one should be able to reduce the size of the reduced integration domain to a minimum, and therefore diminish the costs.
\item The reduced integration domain is updated at each correction of the CH-POD, for a new reduced basis vector needs to be ``observed". This operation is costly as it requires to evaluate quantities on the whole domain. This operation needs to be optimized.
\item The price of the correction is still high. As a result, if the solution which is searched for is far away from the initial snapshot, the method becomes computationally more expensive than a nonlinear direct solver on the complete system. It is thus limited, up-to-now, to the correction of reduced models for the resolution of nearby problems.
In a forthcoming paper, we will propose a multilevel approach in order to perform the correction only in the zones of the structure undergoing the largest topological changes. This strategy should lead to a significant gain in terms of numerical efficiency, and especially in the case of localized nonlinearities, as suggested by the results provided by the patch-assembly procedure.
\end{itemize}

\subsubsection{Extension to finite element problems}

The proposed method has only been validated on damageable lattices so far. Slight differences in the efficiency results might appear when dealing with 2D or 3D softening problems discretised by the finite element method:
\begin{itemize}
\item the conditioning of lattice problems is usually better than the one obtained when using finite element discretisation. We have been able to use a very simple preconditioner for the iterative corrections of the reduced basis: deflation orthogonally to a few Ritz vectors by the SRKS method, and diagonal preconditioner on the resulting deflated problem. Obviously, an enhanced preconditioner should be preferred when dealing with finite element problems, in order to keep the costs of the corrections low.
\item the connectivity in lattice structures can be very high. As a result, the reduced integration domain associated to each of the controlled nodes in the hyperreduction method is quite large, especially compared to the ones obtained when using structured finite element meshes. Therefore, the cost reduction associated to this particular step of the CH-POD should increase when decreasing the connectivity of nodes.
\end{itemize}

One of the major issues to be addressed when dealing with FE softening problems is the question of localization. It is well known that nonlocal damage models should be used to avoid spurious dependency of the solution to the mesh. The efficiency of the hyperreduction technique, which focuses on local patches in the structure, has not yet been investigated in this framework. Its extension does not seem trivial.

\subsubsection{Suitability to parallel computing}

The extension of our strategy to parallel computing is quite straightforward. The H-POD can be implemented in parallel as follows: the domain is divided into subdomains, assigned to separate processors. Solving the small problems resulting from the linearisation of the reduced problem is done systematically on every processor. The evaluation of the internal forces is a local operation performed on the reduced integration domain, distributed on the processors. It only requires an assembly operation, of small numerical costs as involving only degrees of freedom associated to the controlled nodes.

The corrections performed in the CH-POD are very easy to parallelize as they are based on a conjugate gradient algorithm. The projection operation itself has been widely extended to parallel computing (see \cite{gosseletrey2006} for example). In addition, using the condensation method, as proposed in the Schur-complement-based domain decomposition methods would provide better preconditioner than the one used in this work.




\section{Conclusion}

We developed an efficient corrective tool for the adaptive model order reduction of highly nonlinear mechanical problems. The novelty of the approach is that it completely integrates the corrections inside the projection framework classically used in model order reduction. More precisely, the corrections are performed using an iterative conjugate gradient, which is enabled by the strong link between, on the first hand, POD-based model order reduction and, on the other hand, projected Krylov algorithms. The resulting method establishes a bridge model order reduction and classical Newton/Krylov solvers, ensuring its versatility.

The robustness of the strategy has been demonstrated by correcting the reduced model associated with a very complex case of damage propagation. In terms of raw CPU time, our algorithm had been coupled to the hyperreduction method in a straightforward way, making it computationally efficient in the case of mechanical problems involving internal variables.

Yet, the corrections are still expensive compared to solving the initial reduced model. As explained briefly in the last section of the paper, the corrections need not to be done everywhere in the structure, especially when localization phenomena are involved. This can lead to prohibitive costs if the solution that is searched for is distant from the snapshot.
Therefore, a possible enhancement to increase the efficiency of the adaptive model order reduction is to use a multilevel algorithm to perform the corrections.



\section*{Acknowledgments}

St\'ephane Bordas and Pierre Kerfriden would like to thank the support of the Royal Academy
of Engineering and of the Leverhulme Trust for his Senior Research
Fellowship entitled "Towards the next generation surgical simulators"
as well as the support of EPSRC under grants EP/G069352/1 Advanced
discretisation strategies for "atomistic" nano CMOS simulation and
EP/G042705/1 Increased Reliability for Industrially Relevant Automatic
Crack Growth Simulation with the eXtended Finite Element Method.

\bibliographystyle{unsrt}
\bibliography{bibliography}

\end{document}